\documentstyle[12pt]{article}

\begin{document}
\title{Self-consistent treatment of the quark condensate and
hadrons in nuclear matter
}
\author{E.G.Drukarev, M.G.Ryskin, V.A.Sadovnikova}
\date{}
\maketitle
\begin{center}
Petersburg Nuclear Physics Institute, Gatchina, St.Petersburg
 188350, Russia\\
e-mail: drukarev@thd.pnpi.spb.ru, ryskin@thd.pnpi.spb.ru,\\
sadovnik@thd.pnpi.spb.ru
\end{center}
\begin{abstract}
We calculate the contribution of pions to the $\bar qq$-expectation value
$\kappa(\rho)= <M|\bar qq|M>$ in symmetric nuclear matter.  We employ exact
pion propagator renormalized by nucleon-hole and isobar-hole
excitations. Conventional straightforward calculation leads to
the "pion condensation" at unrealistically small values of
densities, causing even earlier restoration of chiral symmetry. This
requires  a self-consistent approach, consisting in using the models,
which include direct dependence of in-medium mass values on
$\kappa(\rho)$, e.g.  the Nambu--Jona-Lasinio--model.  We show, that in
the self-consistent approach the $\rho$-dependence of the condensate is
described by a smooth curve. The "pion condensate " point is removed to
much higher values of density. The chiral restoration does not take
place at least while $\rho<2.8\rho_0$ with $\rho_0$ being the
saturation value.  Validity of our approach is limited by possible
accumulation of heavier baryons (delta isobars) in the ground state of
nuclear matter.  For the value of effective nucleon mass at the
saturation density we found $m^*(\rho_0)=0.6m$, consistent with
nowadays results of other authors.

\vspace{1cm}

 \end{abstract}

\vspace{1.cm}

\section{Introduction}
In this paper we present the calculation of density dependence of the
scalar quark condensate $\kappa=<M|\bar qq|M>$ in  symmetric nuclear
matter. The
 gas approximation \cite{1} provides linear dependence, while account of
interactions in medium leads to nonlinear contributions. Although at normal
density $\rho_0$ the nonlinear corrections are rather small \cite{1}
--- \cite{11}, it is interesting to follow their behaviour while the
density increases.

We show, that the main contribution to nonlinear terms comes from the
interaction of the scalar quark operator with the pion cloud. Thus we must
average the operator $\bar qq$ over in-medium pions.

Concrete calculation are carried out by use of Feynman diagram technique
(Fig. 1) with the pions described by propagators, renormalized by
interactions with medium. The propagation of pions in medium is a special
story, related closely to the problem of pion condensation
 \cite{14}, \cite{23}.  The key point of the latter phenomena is
that a solution of dispersion equation with negative frequency squared,
$\omega_c^2 <0$ emerges at certain value $\rho_c$. This signals the
change in the structure of the ground state. The general belief is that
the phenomena does not take place up to the values of about twice the
normal density.

We show  that the pion condensation is related to the appearance of a new
branch of the solutions of pion dispersion equation on the physical sheet of
pion frequencies, besides the well-known branches with pion quantum numbers
(these are the pion, isobar and sound branches).

Account of the  condensation singularity in the calculation of
$\kappa=<M|\bar qq|M>$ leads to  important physical consequence: the value
of $\kappa$ turns to zero at the density values smaller than $\rho_c$.  Thus, while  the density increases, the
chiral phase transition takes place earlier, than the pion condensation does.

We show that the value of the critical density $\rho_c$ (and thus, the value
of the density of the chiral phase transition) depends strongly on the
 magnitude of the effective nucleon mass $m^*$ in  medium. This happens
because the pion polarization operator $\Pi$ is proportional to $m^*$, thus
 entering the dispersion equation and propagator. The  effective mass $m^*$
  decreases while the density increases. The decrease  of $m^*(\rho)$ causes
the diminution of the polarization operator reflecting the weakening of the
medium influence on the pions. This results in a shift of the point of the
pion condensation to a larger density.

However, assuming any of conventional nuclear physics equations for
in-medium mass $m^*$ with direct dependence on the density, we find the
chiral restoration point to be dangerously close to the saturation value.
This would require strong precursors  of chiral restoration at normal
densities, in sharp contradiction to our knowledge.  Thus, the intermediate
result of our paper is that straightforward application of the pion nuclear
physics to the calculation of the scalar $\bar qq$ expectation value in the
nuclear matter leads to  unphysical results.  The problem  is solved by
self-consistent treatment of the condensate and of hadron parameters.

Indeed, the expectation value of the quark condensate in nuclear matter is
calculated by Feynman diagrams technique. The expressions, corresponding to
the diagrams, include dependence on a number of in-medium hadron parameters.
These are nucleon and pion masses, the coupling constant, the number of
quark-antiquark pairs in pion, etc.  On the other hand, these in-medium
characteristics can be expressed through the averaged values of quark (and
gluon) operators in the framework of QCD sum rules \cite{1}, or by using
other models, e.g. Nambu--Jona-Lasinio--model (NJL). Such models, combined
with the idea of  scaling, developed by G.E.Brown and M.Rho \cite{19},
 \cite{18} enable us to express the baryon effective masses $m^*$,
 $m^*_\Delta$ and pion decay constant $f^*_\pi$, through in-medium value of
 the quark condensate ($m^*(\kappa), f^*_\pi(\kappa)$).

Thus, we solve the following system of equations:

$$\kappa=f_\kappa (m^*,...),$$
$$m^*=f_m(\kappa),$$
$$...$$
with the dots standing for other hadron parameters, depending on $\kappa$.
As a result, there appears a self-consistent scheme for the calculation of
the expectation value of quark condensate $\kappa$ and effective baryon mass
$m^*$ in the medium.
 The
self-consistent calculation leads to a rapid decrease of the effective
nucleon mass with density. Thus, there is no pion
condensation at least up to the density $\rho=2.8\rho_0$.

At larger densities $\rho\geq 2.8\rho_0$
 the heavier baryons (isobars) can be accumulated in
nuclear matter, thus
changing the structure of its ground state. Therefore, the first
phase transition, which takes place while the density increases is
the condensation of heavier baryons in the ground state of
nuclear matter.
If we neglect
the new Fermi sea of isobars, we find that $\kappa$ approaches zero
asymptotically, i.e.  at $\rho\to\infty$.  Also, there is no pion
condensation.

In the self-consistent approach the shape of  $\rho$-dependence of both functions
$m^*(\rho)$ and $\kappa(\rho)$ does not change much, while we modify the
shape of the dependence $m^*(\kappa)$.  Dependence  of the values of $m^*$
 and $\kappa$ on the values of parameters describing effective particle-hole
interactions and on those of form factors of the pion vertices is also weak.

\vspace{0.5cm}{\bf Scalar quark condensate}

Investigation of the scalar condensate can be interesting from several
points of view. It may appear to be useful in the attempts to find the
bridge between description of strong interactions in hadron and quark-gluon
degrees of freedom.  Indeed, the condensate is determined through quark
degrees of freedom, depending, however on the values of hadron
parameters.  On the other hand, it describes the properties of the
matter as a whole.  Being the order parameter of the system, it
characterizes the violation of chiral symmetry.  Its turning to zero
leads to bright consequences for the system as a whole \footnote{On the
other hand, turning  of $\kappa(\rho)=<|\bar qq|>$ to zero may be not
sufficient for the  chiral symmetry restoration.  It is not excluded
that the expectation value$<M|\bar qq|M>$ is equal to zero, but
simultaneously \\ $<M|\bar qq\bar qq\bar qq|M>\neq0$, etc.  However in
many of the models used nowadays, e.g.  NJL-model,  the value $<|\bar
qq|>$ may be considered as an order parameter, and the chiral symmetry
violation does not take place, although  $<|\bar qq|>$ turns to
zero.}.

Also, we hope that the developing of QCD sum rules project started in
\cite{1} will provide the bridge between the two ways of description, based
either on  hadronic or quark degrees of freedom.  By using the QCD sum
rules, the particle properties (mass, pole residue, etc.) can be expressed
through expectation values of the quark and gluon operators.  The scalar
condensate $\bar qq$ is one of the most important. The expectation values of
the vector $\bar q\gamma_0q$ and scalar $\bar qq$ operators play the role of
the vector and scalar boson fields, in terms of the Quantum Hadrodynamics
(QHD) approach \cite{12} in the mean field approximation. In this
approximation the nuclear matter can be considered as a medium with nonzero
expectation value, $<M|\bar q\gamma_0q|M>\neq 0$, and with a new value of
scalar operator $<M|\bar qq|M>$, which is not equal to the vacuum one.

The value of the vector condensate $<M|\bar q\gamma_0q|M>=3<M|\bar
N\gamma_0N|M>=3\rho$ does not depend on the choice of degrees of freedom. The
latter can be quark or hadronic as well. This is due to the vector current
conservation. Thus in the latest equality  $\rho$ should be treated as the
baryon charge density. The  density dependence of the scalar quark
condensate $\kappa(\rho)=\\<M|\bar qq|M>$ is more complicated. It
is the subject of the present paper.

Now we run through the main ideas and results of this paper.  In the gas
approximation \cite{1} \begin{equation}
{\kappa}(\rho)={\kappa}(0)+\rho<N|\bar{q}q|N>,
\end{equation}
where
$\kappa(0) =<0|\bar qq|0> \simeq-0.03$ GeV$^{3}$ is vacuum expectation, and
the number of quarks in a nucleon, $$<N|\bar
qq|N>=\frac{2\sigma}{m_u+m_d},$$ is given by the well known $\pi N$
$\sigma$-term: $\sigma$=45 MeV, $m_u$ and $m_d$ are the current quark
masses.

In a number of papers \cite{1} --- \cite{11} attempts were made to go beyond
the gas approximation

\begin{equation}
 \kappa(\rho)=\kappa(0)+\rho<N|\bar{q}q|N>+S(\rho),
\end{equation}
 with the nonlinear term $S(\rho)$. This contribution
comes mainly from the pion exchange because of small pion mass and
large expectation value
\begin{equation} \eta=<\pi|\bar
qq|\pi>=\frac{2m^2_\pi}{m_u+m_d}.
\end{equation}
Being a Goldstone
meson, the pion describes the collective mode, i.e.  the excitation of
the large number of $\bar qq$-pairs.

The calculations of $S(\rho)$ in one-- and two--loop approximations
(i.e. with one-- and two--pion exchanges) were done in \cite{1,9,11}
\footnote{Note that erroneous isotopic coefficient was used in
\cite{11}.  So, the result for $\varphi_1$ published in \cite{11}
should be multiplied by 3/4.}. At low density the  loop expansion is
equivalent to the expansion with respect to the Fermi momentum
$p_F\propto \rho^{1/3}$.  Each extra loop gives an extra factor
$\rho^{1/3}$(in the case $m_\pi \ll p_F$).

\vspace{0.5cm} {\bf Phase transition}

At large densities, the different types of the phase transitions can
take place  in nuclear matter before the quark-gluon plasma formation.
 There is a number of  possibilities :\\ i) new types of baryons
($\Lambda$, $\Sigma$, $\Delta$) can appear in the ground state of
nuclear matter \cite{13}, when the energy of the nucleon on the Fermi
surface ($\varepsilon _F=p_F^2/2m$) becomes larger than the mass
splitting $\Delta m=m_B-m$ (B=$\Lambda$, $\Sigma$, $\Delta$; $m=m_N$). More
precise condition which takes the baryon-hole interaction and the mass
renormalization into account will be discussed in Sect. 5 and in a separate
paper \cite{15};  \\ ii) chiral invariance will be restored, when
$\kappa(\rho)$ turns to  zero;  \\ iii) the pion condensation \cite{14} can
take place.

Let us clarify the latest point. In  nuclear medium a pion can be
absorbed, producing a free nucleon,  or $\Delta$, and a hole. In
next step the free baryon can emit a new pion and go back filling
up the hole.  These transitions can be interpreted as  the pion
interactions with the particle-hole channels. Thus, instead of
 the free pion in vacuum one deals with the mixture of the pion
and baryon-hole states. These pion-- to baryon-hole transitions
can be  described completely by the polarization operator
$\Pi(\omega,k;\rho)$ in the pion propagator $D$:

\begin{equation} D=\frac{1}{\omega^2-m_\pi^2- k^2
-\Pi(\omega,k;\rho)+i\varepsilon}\, .
\end{equation}
Here and below
$k=|\vec k|.$ The dispersion equation
$$D^{-1}=\omega^2-m^2_\pi-k^2-\Pi(\omega,k;\rho)=0$$ has
several solutions. While the density $\rho$ increases, one of the
solutions, $\omega_c(\rho)$, turns to zero. For the first time it
happens at a critical density $\rho=\rho_c$ for some concrete value
$k=k_c$ of pion momentum.  At larger densities $\rho >\rho_c$, the
square of $\omega_c$ is negative (  $\omega^2_c(\rho)<0$ ) in  some
interval of values of $k$.  In this case one has to add the pion-type
excitations into the ground state of the nuclear matter.  This
phenomenon is called the pion condensation \cite{14}.

Since there is a large number of pions in the ground state, large
contribution to the expectation value $\kappa(\rho)$ appears. Hence,
the value of $\kappa(\rho)$  changes near the point $\rho=\rho_c$
significantly. The pion condensation is the main source of nonlinearity
in the $\kappa(\rho)$ behaviour.

\vspace{0.5cm}

{\bf $<|\bar qq|>$ in the presence of "pion
condensation"}

The aim of this paper is to calculate the quark condensate in the
nuclear medium, with account of possible pion condensation.

We will consider the simplest (one-loop) approximation (Fig. 1) but
with the exact (renormalized) pion propagator (see Eq. (4)) including
geometrical series of baryon-hole insertions.

Of course, this is not the whole set of Feynman graphs, but it
describes and includes all the main physical effects we would like to
discuss.  The short-range interactions will be taken into account in
terms of the Theory of the Finite Fermi System (TFFS) \cite{16}, by
using  the effective constants $g'_{NN}$, $g'_{N\Delta}$,
$g'_{\Delta\Delta}$ corresponding to nucleon and $\Delta$-isobar
rescatterings. On the other hand, the long-range correlations are
described by the exact pion propagator.

Note that in the limit of a small pion-nucleon coupling, our approach
reproduces exactly the one-loop result \cite{1}, \cite{3} and the most
important part of the 2-loop calculations \cite{11}.

If the nucleons are treated in nonrelativistic limit, the value of
polarization operator, accounting the  baryon-hole loop is proportional
to the coupling constant $g^*_A/f^*_\pi$ squared and  to the  nucleon
 effective mass $m^*$:

 \begin{equation} \Pi(\omega,k;\rho)\propto
\left(\frac{g^*_A}{f^*_\pi}\right)^2m^*p_F k^2,
\end{equation}
where
$m^*$, $g_A^*$ and $f^*_\pi$ are the  nucleon mass, axial current and
pion coupling in medium, correspondingly.  Here and below all effective
variables, renormalized in the nuclear medium, are supplied with an
asterisk.  The Fermi momentum $p_F\propto \rho^{1/3}$, and for
symmetric nuclear matter

$$\rho=\frac{4}{(2\pi)^3}\int^{p_F}d^3k.$$ Factor $p_Fm^*$ comes from
the integration of the energy denominator, with $\Delta E\sim
k^2/2m^*$:  $$\int^{p_F}\frac{d^3k}{\Delta E} \sim m^*p_F.$$

Thus, to calculate the true value of quark condensate $\kappa(\rho)$
one has to know the $\rho$-dependence of the baryon mass $m^*(\rho)$
and that of the coupling constant $g^*_A/f^*_\pi$. The simplest
 possibilities are to use either the Landau formula \cite{17}
\begin{equation}
\frac{m^*}{m}=\frac1{1+\frac{2mp_F}{\pi^2}f_1},
\end{equation}
 or the  Walecka-type model \cite{12}, where in
nonrelativistic limit \begin{equation} m^*=m-c\rho
\end{equation}
with certain constant coefficients $f_1$ and $c$.

Using Eqs. (6) or (7), one obtains the  pion propagator  pole at
$\omega=\omega_c(\rho)$ with $\omega^2_c\leq 0$ for the densities
$\rho\geq \rho_c$. The value of critical density $\rho_c$ turns to be
of the order of the saturation one.

Say, the value, obtained in the papers \cite{23}, \cite{34}, is
$\rho_c\sim (1.0-1.5)\rho_0$.  The appearance of  such a pole  was
interpreted as the signal of "pion condensation" \cite{14}.

However, just before the "condensation" (at $\rho < \rho_c$) the
nonlinear contribution $S(\rho)$ increases drastically, and the curve
$\kappa(\rho)$ crosses the zero line. The reason is trivial. When
$\rho\rightarrow \rho_c$, the integral for the pion loop (Fig. 1) takes
the form $$S\sim\int\frac{d\omega
d^3k}{(\omega^2-\omega_c^2(\rho,k))^2}$$ (here we keep the singular
part of integrand only and take into account the symmetry of pion
propagator  in respect to the sign of $\omega$). The integral diverges
near the condensation point $\omega_c(\rho_c,k)\simeq a\cdot(k-k_c)^2$
$\rightarrow 0$ for $k\rightarrow k_c\neq0$:  $$\int \frac{d\omega
k_c^2d(k-k_c)}{ ( \omega^2 - a^2(k-k_c)^4)^2} \rightarrow \infty.$$

This means that one faces another phase transition. Namely, the chiral
symmetry restoration is reached before the pion condensation.  At
larger densities  the pion does not exist any more as a collective
Goldstone degree of freedom, the baryon mass vanishes (if very small
contribution of the current quark masses is neglected), and we have to
stop our calculations based on the selected set of Feynman diagrams (Fig. 1)
with  exact pion propagator.

The structure of pion propagator singularities, various branches  of
solutions of dispersion equation and calculation of the quark
 condensate $\kappa(\rho)$ under the assumptions described by Eqs. (6)
or (7) with the coupling $g^*_A/f^*_\pi$ $=g_A/f_\pi$ $=const$ are
described in Sect. 2, 3, 4.

With  conventional values of TFFS constants ($f_\pi$, $f_1$,
$g'_{NN}$,...) we obtain $\kappa(\rho)=0$ at rather small densities
$\rho\sim (1.1\div 1.2)\rho_0$. This does not look to be realistic.

\vspace{0.5cm} {\bf Self-consistent approach}

Therefore, we consider another approach, which is a self-consistent
 one.  We obtain the expression for quark condensate $\kappa$ which
depends on nuclear density, on effective mass $m^*$, on the effective
constant $f^*_\pi$, etc. On the other hand, the effective
(renormalized) values of $m^*$, $f^*_\pi$~... depend on $\kappa$.

Recall, for example, that in the framework of QCD sum rules the baryon
mass is determined mainly by the $\bar qq$-expectation value. The
relation between the mass and $\kappa(\rho)$ is even more
straightforward in the  NJL-model.  The nucleon (and  constituent
quark) mass is proportional to $\kappa$, and in  medium one finds
\begin{equation} m^*(\rho)=G\cdot \kappa(\rho),
 \end{equation}
where $G$ is the constant for the four-fermion interaction. Thus, in order
to perform self-consistent calculations, we have to solve the  set of
equations

\begin{equation} \kappa=F_\kappa(\rho,m^*(\rho),f^*_\pi(\rho),..)=
\kappa(0) + \rho<N|\bar qq|N>+S(\rho),
\end{equation}
$$\frac{m^*(\rho)}{m}=F_m(\kappa,\rho),$$
$$\frac{f^*_\pi(\rho)}{f_\pi}=F_\pi(\kappa,\rho).$$

In Sect. 5, we use Eqs. (9), together with a hypothesis about
$g^*_A/f^*_\pi$ behaviour, and calculate the expectation
$\kappa(\rho)$.  Two types of behaviour of the ratio $g^*_A/f^*_\pi$
are considered: 1).~ $g^*_A/f^*_\pi$ =$g_A/f_\pi$ =$const$, which is the
latest version of Brown-Rho scaling \cite{18}; and 2). $g_A^*=g_A$;
$f^*_\pi/f_\pi=m^*/m$ \cite{19}.

Here is our main result. For $g^*_A/f^*_\pi$ =$const$ and

$$ \frac{m^*(\rho)}{m}=F_m,\qquad F_m=\frac
{\kappa(\rho)}{\kappa(0)},$$ we get rather smooth $\kappa(\rho)$
dependence. The system tries to prevent the chiral symmetry restoration
at low densities: the ratio $\kappa(\rho)/\kappa(0) \geq 0.2$  up to
$\rho\sim 2.5\rho_0$ ($\kappa( \rho_0)/ \kappa(0)=0.55$).

To study the stability of the results we consider several other
possibilities of   $m^*(\rho)$ dependence.  These are:  $$\frac
{m^*(\rho)}{m}= \left(\frac{\kappa(\rho)} {\kappa(0)}\right)^{1/3} ,$$
and QCD sum rules motivated formula \cite{20}
\begin{equation}
\frac{m^*(\rho)}{m}=\frac{\kappa(\rho)}{\kappa(0)}-
\frac{2.4\rho}{\kappa(0)}.
\end{equation}
with the second term caused by vector condensate.  All the versions
with $g^*_A/f^*_\pi$=$const$ show the same qualitative behaviour. On
the contrary, for $g^*_A=const$ $ $\footnote{As was mentioned in
\cite{19}, the low density evolution from $g_A(0)=1.25$ to
$g_A(\rho_0)=1$ has a special explanation.} and
$f^*_\pi/f_\pi$$=m^*/m$ we reach the critical point $\kappa(\rho)=0$ at
a very small value of Fermi momentum $p_F\simeq 200$ MeV
($\rho<\rho_0/2$), since in this case the polarization operator
increases while $|\kappa|$ decreases ($\Pi\propto m^*/f^{*2}_\pi
\propto 1/\kappa$). The vanishing of the $\bar qq$-expectation
value at $\rho\sim \rho_0/2$ does not take place in the Nature.
Hence, we have to reject this  possibility.

Note that in the present paper we ignore the strange sector, but even
in non-strange sector new baryons ($\Delta$-isobars) appear in the
ground state of nuclear matter at larger densities.  Thus, we cannot
continue calculations at $\rho > (2.5-3.0)\rho_0$, before the
reconstruction of the ground state is carried out.

Since for moderate densities $\rho\sim 2\rho_0$ the effective mass of
nucleon $m^*$ becomes comparable with the Fermi momentum $p_F$, we
account for relativistic kinematics of nucleons.
In this case  we obtained reasonable values of effective
nucleon mass $m^*(\rho_0)=0.6m$ and of the scalar condensate
$\kappa(\rho_0) =0.6\kappa(0)$ for normal nuclear density
$\rho_0$.

\section{The main equations}

The lowest order contribution to the quark condensate beyond the gas
approximation is provided by the nucleon self-energy graph shown in
Fig. 1, with both nucleon and $\Delta$-isobar contributing to the
intermediate state. In  vacuum all intermediate nucleon momenta
$p_2\geq 0$ are available, but in  medium the momenta $p_2>p_F$  are
allowed only, because of Pauli principle (see Fig. 1).  Besides, the
pion in-medium propagator  should be renormalized; this is shown in
Figs. 1b,d by fat wavy line.  Therefore, we calculate the contribution
of the diagrams, shown in Figs. 1b,d subtracting analogous vacuum
 contributions with  bare (vacuum) pion propagator (Figs. 1c,e).

From  formal point of view,  $\kappa(\rho)$ can be calculated as the
derivative of the energy density ${\cal E}$ with respect to the current
quark mass \cite{2}:

\begin{equation}
\kappa(\rho)=\partial {\cal E}/\partial m_q.
\end{equation}
The pion-induced part comes from the differentiation of the nucleon
self-energy $\Sigma_N$, which corresponds to the diagram of Fig. 1a

$$S_\pi(\rho)= \rho\frac{\partial \Sigma_N}{\partial m^2_\pi}
\cdot \frac{\partial m^2_\pi}{\partial m_q},$$
 where the derivative
$$\frac{\partial} {\partial m^2_\pi}
 \frac{1}{( \omega^2 -m^2_\pi- k^2-\Pi)}
=\frac{1}{( \omega^2 -m^2_\pi - k^2-\Pi)^2}$$
squares  pion propagator, and
$$\frac{\partial m^2_\pi}{\partial m_q}=\frac {m^2_\pi}
{m_q}=\eta/2.$$
Here $\eta$ is the number of $\bar qq$-pairs in  pion (Eq. (3)). We  use
linear PCAC equation for the pion mass squared $m^2_\pi$, obtained by
Gell-Mann, Oakes and Renner (GMOR) \cite{21}:

\begin{equation}
m^2_\pi=-\frac{<|\bar qq|>(m_u+m_d)}{2f^2_\pi}.
\end{equation}
In the Feynman graph of Fig. 1, the $\bar qq$ operator is shown by the
fat black point which i)~stands for the pion propagator squared, and
ii)~multiplies the result by the factor $\eta$.

Note that we  average the operator $\bar qq$ over the pion states but
not over intermediate baryon states. Due to the Ward identity, which
corresponds here to the baryon number conservation, all the
contributions containing $<N|\bar qq|N>$ for the $\pi N$ intermediate
state are already accounted for by the second term in Eq. (2).

From technical point of view, this is supported by the following
argument. For the nucleon in the matter the condensate $<N|\bar
qq|N>_m$ can be presented as

$$<N|\bar qq|N>_m=<N_m|\bar qq|N_m>+<N|\bar qq|N>\left(-
\frac{\partial \Sigma_N}{\partial E} \right).$$
Here $|N>$ is free nucleon state and $|N_m>$ is in-medium nucleon one
 with energy $E$; $\Sigma_N$ is the self-energy.  Using the
multiplicative character of renormalization

$$|N_m>=Z^{1/2}|N>,\qquad \frac{\partial\Sigma_N}{\partial E}=Z-1 ,$$
we find $<N|\bar qq|N>_m=<N|\bar qq|N>$, thus proving our assumption.

For  $\Delta$-baryons (Fig. 1), we deal with the contributions
proportional to the difference $$\delta=<\Delta|\bar qq|\Delta>
-<N|\bar qq|N>.$$ Basing on the Additive Quark Model (AQM), we assume
that $\delta=0$.  Thus we take into account  only the pion contribution
 to $\kappa(\rho)$.

Certainly, in the  strong interactions, there is no reasonable parameter
for perturbative series.  We consider the first (one-loop) self-energy
 diagram of Fig. 1  with  full (exact) pion propagator instead. In
other words, we sum up the selected set of Feynman graphs which are
responsible for the lowest singularity in the $\rho$-dependence of
$\kappa(\rho)$. The expression for $S_N(\rho)$, illustrated by
Figs. 1b,1c is:

\begin{equation}
S_N(\rho)=-3\eta
Sp\int\frac{d^3p}{(2\pi)^3}\frac{d\omega d^3k}{(2\pi)^4i}
\end{equation}
$$\left(\Gamma^2_{\pi NN}D^2(\omega,k)\theta (p_F-p)\frac
{\theta(|\vec p-\vec k|-p_F)}{\varepsilon(p)-\omega-\varepsilon(\vec p
-\vec k) +i\delta} \right.$$
$$\left.\,-\Gamma^{02}_{\pi NN}D_0^2(\omega,k)\theta (p_F-p)
\frac{1}{\varepsilon(p)-\omega-\varepsilon(\vec p-\vec k)+
i\delta}\right) .$$
Here $Sp$ stands for summation over the
nucleon spin indices, the factor 3 comes from the summation over
isotopic coefficients and the last term is needed to avoid the
double counting and  subtract the contribution which is already
included into  $<N|\bar qq|N>$ in the second term of Eq. (2),
which is related to the bare nucleon in a vacuum. Recall that
$\rho=2p_F^3/3\pi^2$. We must add similar contribution of
intermediate isobar, illustrated by the diagram of Fig. 1d.

The  free pion propagator is
$$D_0=(\omega^2-k^2-m^2_\pi+i\delta)^{-1}.$$ The $\pi NB$ vertex with B
labeling nucleon or $\Delta$-isobar is
\begin{equation}
\Gamma_{\pi
NB}=\Gamma^{(0)}_{\pi NB} \cdot d_B(k)\cdot x_{\pi NB} ,
\end{equation}
while
\begin{equation} \Gamma^{(0)}_{\pi NN}=\frac{g_A}{\sqrt 2
f_\pi}\bar \psi \gamma_\mu \gamma_5\psi k_\mu = \frac{i g_A}{\sqrt 2
f_\pi} \chi^*(\vec \sigma\vec k)\chi ,
\end{equation}
with $\psi
 (\chi)$ being the (non)relativistic nucleon four- (two-) spinors. In
 order to take into account the nonzero baryon sizes, the bare vertex
$\Gamma^{(0)}$ is multiplied by the form factor taken in a simple pole
form $$ d_B=\frac{1-m^2_\pi/\Lambda_B^2}{1+k^2/\Lambda_B^2}, $$
$\Lambda_N$=0.667 GeV, while  $\Lambda_\Delta$=1.0 GeV  \cite{22}. The
factors $x_{\pi NN}$ and $x_{\pi N\Delta}$ are accounting for the
renormalization of the corresponding vertices due to the particle-hole
pairs. Explicit expressions for them will be given later on (see Eq. (27)).

The expression, corresponding to the diagram of Fig. 1d with
$\Delta$-baryon is analogous to Eq. (13). However, the unrenormalized
 vertex is now

\begin{equation}
\Gamma^{(0)}_{\pi N\Delta} = f_{\Delta/N} \frac{i g_A}
{\sqrt 2 f_\pi}\chi^*(\vec S^+_\alpha\vec k)\chi^\alpha .
\end{equation}
The experiments provide the value of the coupling
constant $f_{\Delta/N} \simeq 2$ \cite{22}, while AQM calculations
give
$f_{\Delta/N} \simeq 1.7$.\\
Certainly, there are no limits for momenta of isobars in the intermediate
states.
Mass difference $\Delta m=m_\Delta-m$ is included into the of
$\Delta$-baryon energy $\varepsilon(\vec p-\vec k)$.

We assume that  the baryon-medium interactions change the potential
energy of any baryon (nucleon or $\Delta$-isobar) by the same value, which
does not depend on the baryon momentum $p$. This is consistent with QHD
picture in the mean field approximation, under the assumption that the
vector field has the same coupling to nucleon and $\Delta$-baryon. In other
words,  AQM is assumed to describe the vector field interaction with baryon.

Note that only  nonrelativistic approximation for the baryon
propagator is used in \\ Sects. 2-5. In  Sect. 6, to estimate
 relativistic effects at large densities (when the Fermi
momentum $p_F$ becomes comparable with effective mass $m^*$) we use
traditional perturbative approach, with all particles on the mass
shell.  Still, we neglect the contribution from badly time-ordered
graphs, where the antibaryon-baryon pairs are created.

Therefore, the only relativistic effect  (Sect. 6) is a relativistic
expression for $\varepsilon(p)$ in the denominator of Eq. (13):

$$\varepsilon_p=\sqrt{(m^{*2} + p^2)} .$$

Now we construct the pion propagator  in nuclear matter. According to
TFFS, we have to sum up the geometric series of baryon-hole loops shown
in Fig. 2a. The contribution of one loop, illustrated by  first
diagrams in Figs. 2b,c are

\begin{equation}
\Pi^{(0)}_N=Sp\int\frac{d^3p}{(2\pi)^3}\Gamma^2_{\pi NN} G_N(\vec
p+\vec k) \theta(|\vec p+\vec k|-p_F)\theta(p_F-p) ,
 \end{equation}
\begin{equation}
\Pi^{(0)}_\Delta=Sp\int\frac{d^3p}{(2\pi)^3}\Gamma^2_{\pi N\Delta}
G_\Delta(\vec p+\vec k) \theta(p_F-p) .
 \end{equation}
 Here the traces are taken over  spin and isospin variables, and
 $G_N$, $G_\Delta$ are  nucleon and $\Delta$-isobar propagators.

In the described  perturbative series for the baryon-hole loop, we
consider both  particle-hole excitation and absorption contributions
(the first and the second diagrams, correspondingly, in  Figs. 2b,c)

 Thus, \begin{equation} \Pi^{(0)}_N=-4\left(\frac{g^*_A}{\sqrt
2f^*_\pi}\right)^2 k^2\left[\Phi_N(\omega,\vec k) +
\Phi_N(-\omega,-\vec k)\right]d^2_N(k),
\end{equation}
\begin{equation}
\Pi^{(0)}_\Delta=-\frac{16}{9}\left(\frac{g^*_A}{\sqrt
2f^*_\pi}\right)^2 f^2_{\Delta/N}
k^2\left[\Phi_\Delta(\omega,\vec k) + \Phi_\Delta(-\omega,-\vec
k)\right]d^2_\Delta(k),
 \end{equation}
 with Migdal's function
\begin{equation}
\Phi_\Delta(\omega,k) =\frac{1}{4\pi^2} \frac{m^{*3}}{k^3}
\left[\frac{a^2-b^2}{2}
\ln(\frac{a+b}{a-b}) - ab\right] .
\end{equation}
Here $a=\omega-k^2/2m^*-\Delta m$, $b=kp_F/m^*$, and
$Re(\Delta m)= m_\Delta - m$, $Im(\Delta m) = -\Gamma_\Delta/2$.
 $\Gamma_\Delta$  is the isobar width.

 Integration over the momenta  $p$ provides

\begin{equation}
\Phi_N(\omega,k)=\frac{m^*}{k}\frac 1{4\pi^2}\left(\frac{-\omega
m^*+kp_F}{2} +\frac{(kp_F)^2-(\omega m^*
-k^2/2)^2}{2k^2}\ln(\frac{\omega m^*-kp_F-k^2/2}{\omega m^*
-kp_F+k^2/2})\right.
\end{equation}
$$\left.-\omega m^*\ln(\frac{\omega m^*}{\omega
m^*-kp_F+k^2/2})\right),$$\\
at $0\leq k\leq 2p_F$, while

\begin{equation}
\Phi_N(\omega,k)=\frac{m^*}{k}\frac
1{4\pi^2}\left(-\frac{p_F}{k}(\omega
 m^*-k^2/2) +\frac{(kp_F)^2-(\omega m^* -k^2/2)^2}{2k^2}
\ln(\frac{\omega
m^*-kp_F-k^2/2}{\omega m^* +kp_F-k^2/2})\right)\\
\end{equation}
at $2p_F\leq k\leq \infty$.

Note that Eq. (22) differs from analogous equation of Refs. \cite{23},
\cite{22} at $k\leq 2p_F$. The short-range correlations originated from the
baryon-hole rescattering are described in terms of TFFS with the help of
effective constants $g'_{NN}$, $g'_{N\Delta}$, $g'_{\Delta\Delta}$, which
correspond to N-N, N-$\Delta$ and $\Delta$-$\Delta$ rescatterings.  If it is
not specially mentioned, we use $g'_{NN}$=1.0, $g'_{N\Delta}$=0.2,
$g'_{\Delta\Delta}$=0.8 \cite{23}.  After summation of the geometrical
series of baryon-hole loops, we obtain the polarization operator
\cite{23}, \cite{24}

$$\Pi(\omega,k;\rho)=\Pi_N +\Pi_\Delta, $$
with
\begin{equation}
\Pi_N=\Pi_N^0(1+(\gamma_\Delta - \gamma_{\Delta \Delta})
\frac{\Pi_\Delta^0}{k^2})/E,
\end{equation}
\begin{equation}
\Pi_\Delta=\Pi_\Delta^0(1+(\gamma_\Delta - \gamma_{NN})
\frac{\Pi_N^0}{k^2})/E.
\end{equation}
Denominator E has the form
\begin{equation}
E=1-\gamma_{NN} \frac{\Pi^0_N}{k^2}-\gamma_{\Delta \Delta}
\frac{\Pi^0_{\Delta}}{k^2} +
(\gamma_{NN}\gamma_{\Delta
\Delta} - \gamma^2_\Delta)\frac{\Pi^0_{NN}\Pi^0_{\Delta}}{k^4}.
\end{equation}
The effective constants $\gamma$ are related to $g'$, as follows:

$$ \gamma_{NN}=C_0g'_{NN}\left(\frac{\sqrt2f^*_\pi} {g^*_A}\right)^2,
 \gamma_\Delta=\frac{C_0g'_{N\Delta}}{f_{\Delta/N}}
\left(\frac{\sqrt2f^*_\pi} {g^*_A}\right)^2,
 \gamma_{\Delta\Delta}=\frac{C_0g'_{\Delta\Delta}}{f_{\Delta/N}^2}
\left(\frac{\sqrt2f^*_\pi} {g^*_A}\right)^2,$$
where $C_0$ is the normalization factor for the effective
particle-hole interaction in the nuclear matter \cite{23}
$$C_0=\frac{\pi^2}{p_Fm^*}.$$

The vertex renormalization factors  introduced in Eq. (14)
 are
 \begin{equation}
  x_{\pi NN}=(1 + (\gamma_\Delta -
 \gamma_{\Delta\Delta}) \frac{\Pi^0_\Delta}{k^2})/E,\qquad
x_{\pi N\Delta}=
(1 + (\gamma_\Delta -\gamma_{NN}) \frac{\Pi^0_N}{k^2})/E.
\end{equation}

\section{Singularities of the pion propagator }

We start with the cuts corresponding
to singularities of the polarization operator $\Pi$. In the
complex $\omega$-plane, the nucleon-hole state induces
logarithmic cuts on the real axis.  As one can see from Eqs. (22-23),
$\Pi^0_N$ has two cuts at positive $\omega$ in the interval $0\leq k\leq
2p_F$

\begin{equation} 1)\quad 0\leq \omega
\leq \frac{kp_F}{m}-\frac{k^2}{2m},\\
\end{equation}
\begin{equation}
2)\quad \frac{kp_F}{m}-\frac{k^2}{2m}\leq \omega \leq
\frac{kp_F}{m}+\frac{k^2}{2m},
\end{equation}
while for large $k\geq 2p_F$ we have one cut only:
\begin{equation}
 -\frac{kp_F}{m}+\frac{k^2}{2m}\leq \omega \leq
\frac{kp_F}{m}+\frac{k^2}{2m}.
\end{equation}
The cut caused by the $\Delta$-hole state lays  below
 the real axis (for $Re$ $\omega>0$) at
\begin{equation}
\frac {k^2}{2m} +\Delta m -\frac{kp_F}{m}\leq \omega\leq
\frac {k^2}{2m} +\Delta m +\frac{kp_F}{m},\\
\end{equation}
with $Im$ $\omega=-\Gamma_\Delta/2$.

Besides, there are symmetric cuts at  $Re$ $\omega<0$.
 The
complete structure of the cuts  of polarization operator is shown
in Fig. 3.

To make the picture more visual we start with fixed
$\Gamma_\Delta$=115 MeV equal to the width of $\Delta$-isobar in
vacuum. On the other hand, the values of $\omega$ which are important
in our integrals are rather small, and practically there is no phase
space for the decay $\Delta\rightarrow \pi N$ in the medium.  Thus,
the $\Delta$-isobar width  in the medium $\Gamma^*_\Delta\simeq 0$.
Therefore, in final computations we  put $\Gamma^*_\Delta = 0$.

Another set of singularities is provided by the poles of the
total pion propagator, i.e. by solutions of
  the dispersion equation
\begin{equation}
D^{-1}=\omega^2-k^2-m^{*2}_\pi-\Pi(\omega,k;\rho)=0.
 \end{equation}
$m^*_\pi$ denotes in-medium value of the pion mass, which is equal to
$$m^{*2}_ \pi=m^2_\pi+\Pi_s$$
with the scalar polarization operator $\Pi_s$ describing S-wave pion-nucleon
rescattering. The polarization operator, provided by Eqs. (24), (25),
accounts for the P-waves only.  For the analysis, carried out in this
Section it is sufficient to use the gas approximation equation
$$\Pi_s=-\rho\frac{<N|\bar qq|N>(m_u+m_d)}{2f^2_\pi}.$$
while in later analysis we use GMOR expression (12) for $m^*_\pi$ and modify
the value of ${f_\pi}$ as well.

 Starting analysis of  Eq. (32) from small values of  densities
($\rho\leq \rho_0$) we find three  branches of its solution on the
 physical sheet.

\subsection{The pion branch $\omega_\pi(k)$}

 At $k\rightarrow 0$ it starts  at $\omega=\pm \sqrt{m^{*2}_\pi}$
leaving the physical sheet through  the  cut of $\Pi^0_\Delta$. For example,
this takes place at $k\simeq 4m_\pi$, when $p_F$=290 MeV. Hence, at large
momenta one deals with the $\Delta$-hole excitations instead of pion's ones.

\subsection{The sound branch $\omega_s(k)$}

It is a slightly changed solution of the equation $E=0$
(see Eq. (26)). At very small momenta $k$ the admixture of other
branches is tiny and the change is negligible. As it should be for
the sound wave, $\omega_s =const\cdot k$ for small $k$. At
$k\simeq 0.43m_\pi$ (at $p_F$=290 MeV) this branch leaves for the lower
sheet through the second  cut of $\Pi^0_N$  (see Eq. (29)).  At $k>
0.43m_\pi$ this solution is  on the second unphysical sheet of the complex
$\omega$-plane.

\subsection{The isobar branch $\omega_\Delta(k)$}

It is mainly the $\Delta$-hole sound wave.  Starting at
$Re(\omega_ \Delta (k=0)) =m_\Delta-m$, $Im(\omega_ \Delta
(k=0)) = -\Gamma_\Delta /2$, it plunges under the isobar
$\Pi^0_\Delta$ cut (at $k=3.8m_\pi$, when $p_F$=290 MeV).

\subsection{The "condensate" branch $\omega_c(k)$}

 This  solution comes
to the physical sheet through the first cut of $\Pi^0_N$  at
$p_F> 283$ MeV/c. The "trajectories" of the solution are shown
in Fig. 4 for different values of $p_F$ and of $\Gamma_\Delta$. The
part which is in the upper half-plane ($Im$ $\omega >0$)
corresponds to the unphysical sheet. For small $p_F\leq 283$ MeV
all the trajectory is placed on the unphysical sheet,  but
at larger $p_F$ it comes down to the physical sheet. Say,  for $p_F$=300
MeV (360 MeV) the solution is on the physical sheet at $k/m_\pi =1.06\div
2.60$ ($k/m_\pi =0.36\div 3.91$). This is illustrated by Fig. 4b.
In Fig. 4a one can see that the real part of $\omega_c(k)$ decreases with
$\Gamma_\Delta$ tending to zero, when $\omega_c(k)$ is  on the physical
sheet. For  $\Gamma_\Delta =0$ the solution goes along the negative
imaginary axis.  Of course, here $\omega_c^2(k)<0$. Thus  this is  the
 singularity responsible for the so-called "pion condensation" \cite{14}.

We have several reasons to put the latest words in the quotation marks. To
start with,  this is not the pion branch ($\omega_\pi$), but another
one.  It starts  at $k=0$ at the same point as the pion branch
$\omega_\pi$ does :  $\omega_\pi(k=0)=m^*_\pi$. However, it goes to the
other  sheet. Note also, that  while $Re$~ $\omega_c>0$, the imaginary
part of the solution is negative everywhere on the physical sheet. This
means that  we never face the mode with $Im$ $\omega_i>0$. In other
words, there is no "accumulation of pions", contrary to the naive
 understanding of condensation.

Of course, there are  singularities with $Im$ $\omega_i > 0$
in the left half-plane of $\omega$. However, these singularities
originate from the inverse time-ordered graphs of Fig. 2c.
They are caused by the terms $\Phi (-\omega, -\vec k$) in Eqs. (19),
 (20).  These singularities  correspond to "antiparticles" and do not
describe solutions which grow with time.

Nevertheless, the fact that even for $\Gamma_\Delta =0$ one obtains a
nonzero imaginary part $Im$~ $\omega_c\neq0$ and
$\omega^2_c<0$ signals on certain instability of the solution. When
$\omega_i^2$ turns to zero (for any $k=k_c$) at certain
$\rho= \rho_c$, the ground state should be reconstructed. New
components, like baryon-hole excitations (with the pion
quantum numbers)  emerge in the ground state of
nuclear matter. Thus, we cannot use the
same approach  at larger values of $\rho > \rho_c$.

Thus, the appearance of the  singularity
$\omega^2_c= 0$ on the physical sheet
shows, that  phase transition takes place in the nuclear
matter.

We have to emphasize that in all  calculations here and below
 we  use the Landau effective mass (6) with the
coefficient $f_1$, which gives $m^*(\rho_0)= 0.8m$; then $g^*_A
=1.0$ and $f^*_\pi= f_\pi= 92$~MeV. The mass splitting $\Delta m
=m_\Delta^* -m^* =const$ with $Re\Delta =292$ MeV.  The values
of TFFS effective constants are $g'_{NN}=1.0$,
$g'_{N\Delta}=0.2$, $g'_{\Delta\Delta}=0.8$. The constant in
the $\pi N\Delta$-vertex (16) is $f_{\Delta/N} =2$.

The dependence of the concrete values of $\omega_i(k)$ on the values of TFFS
constants $g'_{NN}$, $g'_{N\Delta}$, and $g'_{\Delta\Delta}$, being changed
in reasonable limits, is weak. We can say the same about the
dependence on the value of the coupling constant $f_{\Delta/N}$.
The whole picture is more sensitive to the values of effective mass $m^*$
and to that pion-baryon coupling $g^*_A/f^*_\pi$. This will be the subject
 of the analysis carried out in next Sections.

\section{Pion contribution to the quark condensate}

In order to carry out integration over $\omega$ in the integral in
right hand side of Eq. (13), we specify the integration contour in
complex plane.  The contour should go below  the pion propagator
singularities in the left half-plane $\omega$ and above the
singularities in the right one.  We have chosen a straight line $Im$ $
\omega =a\cdot Re$ $ \omega$ with a slope $a\leq 1$. Since the Cauchy
integral is convergent, the result of integration does not depend on
the slope value  $a$, that is proved by our computations.

The results of calculation of the function  $S(\rho)$, defined
by Eqs. (2), (13) are shown in Fig.~ 5.
 For the sake of convenience, we display the ratio

$$\frac{\kappa(\rho)}{|\kappa(0)|}=-1 +
\rho\frac{<N|\bar qq|N>} {|\kappa(0)|} +
\frac{S(\rho)}{|\kappa(0)|}.\eqno{(2.1)}$$
The most interesting  events take place at
 $p_F$ between 270 MeV and 320 MeV. The large change of the values
of $S(\rho)$ is
due to the "pion condensation" singularity
$\omega_c$,  coming to the physical sheet
very close to the integration contour
at $p_F= 283$ MeV.

Fig. 5a illustrates the weak dependence of the
behaviour of the function $S(\rho)$ on the values of TFFS parameters and
on that of  $f_{\Delta/N}$.

The value of the width of $\Delta$-isobar is much more important. At
smaller $\Gamma_ \Delta$ the resonance-like structure becomes more
pronounced (see Fig. 5b). The peak becomes higher and more narrow. At
$\Gamma _\Delta \rightarrow 0$ the poles at $\omega =\omega_c (k)$
pinch the contour at $\omega_c(k_c) =0$, $k_c\neq0$ leading to the
infinite value of $S(\rho)$, when the "pion condensate" singularity
 emerges for the first time on the physical sheet at $\omega_c=0$. Recall
that namely zero value of $\Gamma _\Delta$ is expected in nuclear medium for
a small $\omega$.

However, as is was discussed in the Introduction, before the pole at
$\omega =\omega_c$ (at $\rho =\rho_c$)  reaches the physical sheet,
$S(\rho)$ becomes so large that it cancels the negative vacuum expectation
value $\kappa(0) \simeq -0.03$ GeV$^{3}$, and the whole scalar quark
condensate turns to zero.  The vanishing of the scalar quark condensate
indicates  the chiral invariance restoration, and after that one has to deal
with quite another system, i.e. with another phase of the nuclear matter.

The prediction of  chiral phase transition at rather low values of
$\rho$ (close to normal nuclear density $\rho_0$)  looks too strong.
On the other hand, the results are stable enough and do not change too
much under the variation of TFFS parameters and that of
$f_{\Delta/N}$. These statements are true for both types of behaviour
of nucleon effective masses.  However, the dependence on the numerical
values of the coefficient which enter Eqs. (6), (7) is strong. The
coefficient can be fixed by the choice of the value of $m^*(\rho_0)$.
 The results for $m^*(\rho_0)=0.8m$  and for $m^*(\rho_0)=0.7m$ are
 compared in Fig. 5c.  In the latter case, for the smaller effective
mass, the phase transition takes place at $p_F\simeq 320$ MeV, i.e. at
larger values of density.

\section{Self-consistent approach}

\subsection{Assumptions on the density behaviour of hadron parameters}

As we showed  above, the phase transition
(either "pion condensation" or  chiral symmetry restoration)
density value depends strongly on the value of baryon effective mass
$m^*$. On the other hand, in the framework of commonly used
models the
hadron mass depends mainly on the scalar quark condensate
$\kappa(\rho)$.  This problem should be solved self-consistently.

Of course, it would be nice to calculate all the masses ($m^*$,
$m^*_\Delta$, $m^*_\pi$) and constants ($g^*_A$, $f^*_\pi$,..) with
the
help of QCD sum rules, substituting them in the next step
into our expression for
$\kappa(\rho)$,  solving the equation
\begin{equation}
\kappa=\kappa(\rho,m^*(\rho,\kappa),f^*_\pi(\rho,\kappa),...)
\end{equation}
in  the final step.  Unfortunately, it is not so easy. One of the main
obstacles is that the  masses and hadron constants depend not  on the value
of the scalar quark condensate only but on the in-medium expectation values
of other operators (usually, more complicated ones) as well.

Therefore,  we use simplified  scenario of  the $\kappa$ dependence of the
mass $m^*$ and of the other parameters involved.

Fortunately, there are model-independent equations. The GMOR relation can
be generalized to the case of finite density due to PCAC:
\begin{equation}
m^{*2}_\pi=-\frac{\kappa(\rho)}{2f^{*2}_\pi}(m_u+m_d).
\end{equation}
Also, due to chirality, the number of quark-antiquark pairs
inside  pion is

\begin{equation}
 \eta^*=\frac{2m^{*2}_\pi}{(m_u+m_d)}.
\end{equation}
Now we come to model-dependent relations. Denote the ratios
\begin{equation}
\alpha(\rho)=\frac{m^*}{m},\quad \beta(\rho)=\frac{f_\pi}{f_\pi^*}.
\end{equation}
In the NJL-model \cite{19},

\begin{equation}
\alpha(\rho)=\frac{\kappa(\rho)}{\kappa(0)}.
\end{equation}
In order to check the stability of our result, we perform the
calculations using the  two more types of $\kappa$-dependence of
$\alpha$ :

\begin{equation}
\alpha(\rho)=\left(\frac{\kappa(\rho)}{\kappa(0)}\right)^{1/3}.
\end{equation}
This latter expression can be justified by the dimensional counting, if
there is only one dimensional parameter. Unfortunately,
 in our case, this argument does not work, since there are
 at least two external
dimensional parameters $\rho$ and $\Lambda_{QCD}$,

Another expression for $\alpha(\rho)$ is motivated by the
 QCD sum rules analysis \cite{20}:
  \begin{equation}
\alpha(\rho)=\frac{\kappa(\rho)}{\kappa(0)}-
\frac{2.4\rho}{\kappa(0)}
\end{equation}
with the last term in the right hand side caused by the vector condensate.

Now we discuss the in-medium behaviour of pion
decay constant $f^*_\pi$, which is
proportional to the  pion radius inverted, i.e. $f_\pi \propto \sqrt{N_c} /
r_\pi$ \cite{ann118}.
In the NJL-model near the phase transition point
($\kappa \rightarrow 0$), the pion radius  increases
unlimitedly, and $f_\pi \rightarrow 0$ when $\kappa =<|\bar qq|>
\rightarrow 0$.  Thus, it looks natural to assume that

$$\frac{f^*_\pi}{f_\pi}=\frac{m^*}{m}.$$

Brown and Rho \cite{19} made even stronger hypothesis assuming
 that all the parameters of the same
dimension are proportional to each other
\begin{equation}
\frac{m^*}{m}= \frac {m^*_\Delta} {m_\Delta}
=\frac{f^*_\pi}{f_\pi}...= \alpha(\rho)\mbox{,   i.e. }
\beta=\frac 1{\alpha},
 \end{equation}
while the dimensionless parameters do not change in nuclear
medium; in particular, (see footnote 3)
  $$g_A^* =g_A=const.$$

The second, alternative  hypothesis is based on the idea of
confinement.  If the deconfinement phase transition does not take
place
simultaneously with the chiral invariance restoration, the pion
radius
should be limited and $f^*_\pi$ has a nonzero value when we approach
the chiral transition point. Therefore, we  consider below another
limiting possibility:

\begin{equation}
f^*_\pi =f_\pi\mbox{,   i.e. }  \beta = 1.
\end{equation}
In  the present calculations  we have fixed the value
of the axial coupling constant, $g_A^*=~1$. The most important  parameter in
our calculation is the ratio $g^*_A/f^*_\pi$ (see Eq. (5), (19), (20)).
Thus, just as in the latest version of Brown-Rho scaling \cite{18},  $g^*_A/
f^*_\pi =const$, i.e. does not depend on $\rho$.

Note that  both hypotheses, described by Eqs. (40), (41) are consistent
 with the QCD sum rules for pion \cite{25}, which can be generalized
for the case of finite density in a straightforward way  (the proof
will be published elsewhere):

\begin{equation}
  \frac{\pi} {2}\left( \frac{f_\pi^*
m^{*2}_\pi} {m_u+m_d}\right)^2= \frac{3W^{*4}_0} {32\pi}\left(
\frac{\alpha_s(W^{*2}_0)} {\alpha_s(\mu)} \right)^{8/b} +
\frac{\pi}{16} <\frac{\alpha_s} {\pi}G^2_{\mu\nu}> .
\end{equation}
Here $W^*_0$ is the continuous threshold value, i.e. the minimal
energy of the multihadronic states with  pion quantum numbers;
$\alpha_s$ is the QCD coupling and $G^2_{\mu\nu}$ is the gluon field
squared.

Neglecting the last (numerically small) term and anomalous
dimension ( i.e. putting $\alpha_s(W_0^2)$ =$\alpha_s(\mu)$), one can
satisfy Eq. (42) in two ways:  (i) $f^*_\pi = f_\pi= const$
and the threshold position $W^{*2}_0 \propto m^{*2}_\pi$, or
(ii) the fixed threshold $W^*_0 =W_0 =const \sim 1$ GeV and
 Eq. (40) is consistent with for $f^*_\pi \propto \kappa (\rho)$

We must also make assumptions on  in-medium value
of the $\Delta$-isobar mass. If  Eq. (40) is true, the
$\Delta$-isobar--nucleon mass splitting satisfies the relation
$$ \frac
{(m^*_\Delta-m^*)}{(m_\Delta-m)} =\frac{\Delta m^*}{\Delta m}
=\alpha(\rho) =\frac {1}{\beta}.$$
However, if  $\beta=1$ (Eq. (41)) we come to  $\Delta m^*=\Delta m$.

The experimental situation with  $\Delta$-isobar mass in nuclear matter is
not quite clear at the moment. On the one hand, the total photon-nucleus
 cross section indicates that the mass $m^*_\Delta$ does not decrease in the
medium \cite{26}, while the nucleon mass $m^*(\rho)$ diminishes with
$\rho$.  This means that the splitting $\Delta m^*$ increases ($\Delta
m^* > \Delta m$), opposite to the  $\rho-\pi$ mass splitting. This fact
(if it does take place) looks strange, since the two kinds of splitting
are caused by the same colour magnetic (spin-spin) quark-quark
interaction. On the other hand, the experimental data for total
pion-nucleus cross sections \cite{27} are consistent with the mass
$m^*_\Delta$ decreasing in the matter. As to calculations, the
 description within the Skirmion model \cite{28}  predicts that
$m^*_\Delta$ decreases in nuclear matter and $\Delta m^* < \Delta m$.
Equation $\Delta m^* = \Delta m$ is also true in Walecka model, if AQM
prediction  for the scalar field-baryon coupling $g_{sNN} =g_{s\Delta
\Delta}$ is assumed.

Thus we expect that the  phenomenological parametrization,
$$\frac{\Delta m^*}{\Delta m}=\frac {1}{\beta (\rho)},$$
i.e. $\Delta m^* = \Delta m$ when $\beta =1$ does not look too unlike.

\subsection{A rejected scenario}

Here we try the scaling, provided by Eq. (40), with $\alpha(\rho)$ given
by (37) and $\beta =1/\alpha$.  We find  the phase transition to take
place at the densities, about 2.5 times smaller than the normal one,
($p_F \simeq 200$ MeV).

The technical  explanation is simple. Polarization operator given by Eqs.
(5,19,20,24-26), which is responsible for the  "pion condensation"
singularity $\omega_c(k)$, behaves as

$$\Pi\propto\frac {m^*}{f^{*2}_\pi}\propto\frac 1{\kappa(\rho)}
\propto
\frac 1{\alpha(\rho)}.$$

While the density $\rho$ increases, the value of
$|\kappa(\rho)|$ and that of $\alpha(\rho)$ become smaller. Polarization
operator increases and the "pion condensate" singularity $\omega_c$
approaches the physical sheet at smaller $p_F$.  The nonlinear pion
contribution $S(\rho)$ becomes very large and the whole value $\kappa(\rho)$
tends to zero (dashed curves in Fig. 7). Also, since $\Pi\propto
1/\alpha(\rho) \rightarrow \infty$ when the value of $\kappa(\rho)$ turns to
zero, we lose the solution of equation (33) for $\kappa(\rho)$  (see
Appendix for details), which from now on becomes the complex one.

Since that, one has to deal with another phase of the
matter, with  much smaller particle masses. The whole picture
contradicts sharply to our knowledge about nuclei and nuclear matter. Thus,
we reject this scenario.

\subsection{ An accepted scenario }

For $\beta =1$ (Eq. (41)) the situation looks much better. Up to
$\rho
\simeq 2.5\rho_0$, we deal with a self-consistent  solution of the
system (9).  The value of quark condensate tends to zero, when that of the
density $\rho$ increases, and we never reach either the chiral
symmetry restoration or "pion condensation".

Self-consistent set of Eqs. (9), is solved numerically using
the standard iteration procedure. Technically this means, that calculation
of $\kappa(\rho)$ with vacuum parameters is followed by calculation of
$\alpha(\rho)$. In the next step we obtain in-medium values of all the other
parameters by using Eq. (40). This enables us to obtain $\Pi_N$ and
$\Pi_\Delta$ provided by Eqs. (24)-(27). In the last of the cycle
substitution of these operators into Eq. (13) for further integration
provides a new value of $\kappa(\rho)$. Thus we come to a new cycle.

The validity of our calculations for the larger values of $\rho$ is
 limited by another phase transition. At some value of Fermi momentum
($p_{F\Delta} \propto \rho_\Delta^{1/3}$), the {\em total} energy of a
nucleon on Fermi surface becomes larger than the energy of the
$\Delta$-isobar at rest, i.e. $E_\Delta(0) \leq E_N(p_F)$.  Thus, the
isobars starts to be accumulated by the ground state of nuclear matter.
This effect, as well as possible appearance of other types of baryons,
can be taken into account in our scheme.  Thus our calculations are
reliable below $p_F$, corresponding to this phase transition and mark
the corresponding points by the black circles on the curves shown in
Figs.  6, 7.

From the technical point of view, the effect of instability with
respect to the $\Delta$-isobar accumulation in the ground state reveals
itself in the fact that  the branching points (left edge of
the right $\Delta$-hole cut (with $Im$ $\omega<0$) and right edge of the
left $\Delta$-hole cut (see Fig. 3)) cross the vertical axis $Im$
$\omega$=0.  And two $\Delta$-hole cuts start to deform, i.e.  to cross or
to pinch (for $\Gamma_\Delta$=0) the integration contour in the
$\omega$-plane (Eq. (13)).

The possibility for a other, but nucleons, types of baryons to be
contained in the nuclear-matter ground state was discussed in
\cite{13}. The case of $\Delta$-isobar in the presence of a pion
condensate was considered in \cite{29}, \cite{30}; the problem without the
$\pi$-condensation was studied in terms of Walecka model in
\cite{31}, \cite{32}.

The results plotted in Fig. 6a (in terms of $\kappa(\rho)$) and Fig. 6b
(in terms of $m^*(\rho)$) do not change too much under reasonable
 variations of the TFFS couplings.  Instead of $g'_{NN}$=1.0 in the
master version (solid curve), in Fig. 6 we put $g'_{NN}$=0.7
(dot-dashed curve); for dotted curve $g'_{\Delta \Delta}$=1.2 instead
of 0.8; for dashed curve $f_{\Delta/N}$=1.7 instead of 2.0. In all
calculations in this Section we have used $\Gamma_\Delta$=0.

The lower solid line in Fig. 6b (and Fig. 7b) is drawn for the limit
values of $m^*/m$ at every value $p_F$. If  the ratio $m^*/m$ is smaller
than the limit value, the ground state of nuclear matter contains
 isobars.  The equation for this line is determined by the condition
that the isobar logarithmic cut (Eq. (31)) of the polarization
operator $\Pi^0_\Delta$ (Eq. (20)) touches  the vertical axis $Im$
$\omega=0$ see (Fig. 3).

In Fig. 7 we demonstrate the dependence of the value of the scalar
condensate on the type of the  scaling function $\alpha(\rho)$. Solid
curve  corresponds to the master version, with $\alpha(\rho)$ given by
Eq. (37) and $\beta=1$.  Two other parametrizations, provided by
Eqs. (38), (39) shown by dotted and dot-dashed curves. The results,
corresponding to the law, described by Eq. (38), differ quantitatively from
the two others.  However, as we said earlier, the latter are better based.

\section{Account of relativistic kinematics}

At moderate densities of about twice the normal value, the value of
effective mass $m^*$ becomes comparable with that of  Fermi momentum. Thus,
one cannot neglect the relativistic effects any more. We take into account
relativistic kinematics by using relativistic expression for the energies
$\varepsilon_k= \sqrt{(m^{*2}+k^2)}$ in all the formulae.  However, we still
 omit the baryon-antibaryon pair contributions.

In terms of traditional perturbative theory, Eqs. (21)-(23) should be
replaced now by:

\begin{equation}
\Phi(\omega,k) = \Phi^{(1)}_N(\omega,k) \theta(p_F-k) +
\Phi^{(2)}_N(\omega, k) \theta (2p_F-k)\theta(k-p_F) +
\Phi^{(3)}_N(\omega, k) \theta(k-2p_F);
\end{equation}

$$
\Phi^{(1)}_N(\omega,k) =\int^{p_F}_{p_F-k}dpA(\omega,k),\quad
\Phi^{(2)}_N(\omega,k) =\int^{p_F}_{k-p_F}dpA(\omega,k)+
\int^{k-p_F}_{0}dpB(\omega,k),
$$
$$\Phi^{(3)}_N(\omega,k) =\int^{p_F}_{0}dpB(\omega,k) ,$$
where
\begin{equation}
A(\omega,k)=\frac{p}{4\pi^2}\frac{m^*}{k} \frac
{m^*}{\sqrt{(p^2+m^{*2})}}
\ln\left(\frac{\sqrt{(p_F^2+m^{*2})} -\omega-\sqrt{(p^2+m^{*2})}}
{\sqrt{((p+k)^2+m^{*2})} -\omega-\sqrt{(p^2+m^{*2})}}\right) ,
\end{equation}
\begin{equation}
B(\omega,k)=\frac{p}{4\pi^2}\frac{m^*}{k} \frac
{m^*}{\sqrt{(p^2+m^{*2})}} \ln\left(\frac{\sqrt{((p-k)^2+m^{*2})}
-\omega-\sqrt{(p^2+m^{*2})}} {\sqrt{((p+k)^2+m^{*2})}
-\omega-\sqrt{(p^2+m^{*2})}}\right) ;
\end{equation}
\begin{equation}
\Phi_\Delta(\omega,k) =\int^{p_F}_{0}\frac {p}{4\pi^2}
\frac{m^*}{k}
\frac{(m^*+m^*_\Delta)}{2\sqrt{(p^2+m^{*2})}} \ln\left(\frac{\sqrt
{((p-k)^2+m^{*2}_\Delta)} -\omega-\sqrt{(p^2+m^{*2})}} {\sqrt
{((p+k)^2+m^{*2}_\Delta)} -\omega-\sqrt{(p^2+m^{*2})}}\right) .
\end{equation}
By using these relativistic expression, we can extend the
self-consistent calculation up to $\rho \simeq 2.8\rho_0$.  The limit
is still determined by  transition to the isobar accumulation phase. The
results of calculation of the scalar condensate for the three
considered possibilities of dependence of the effective mass on
$\kappa$ are presented in Fig. 8.  In the  QCD sum rules motivated
 parametrization, described by Eq. (39) with $\beta =1$, we find
 $m^*/m=0.67$, at normal nuclear density $\rho=\rho_0$. And we have
$m^*/m=0.6$ in parametrization (37) with $\beta=1$. The value is in good
agreement with the one, obtained recently in framework of QHD \cite{33}.

\section{Summary}

Since the  gas approximation equation for the quark condensate
$\kappa(\rho)$ was presented in \cite{1}, the nonlinear contribution
$S(\rho)$ was considered in a number of papers. The analysis of M.Ericson et
 al.  \cite{7}, \cite{8} was based on the general properties of the $\pi N$
scattering amplitude and its generalization  for the case of nuclear medium.
As usually, some uncertainties come from the fact that one deals with the
off-mass-shell amplitude.  Thus, certain assumptions about the
$NN$-interaction and on $\rho$-dependence of the effective pion mass
$m^*_\pi$ in medium are needed.

Another group of papers \cite{3}, \cite{4}, \cite{8} was based on NJL-model.
However, in this case they discussed not the nuclear (build up of the
hadrons) medium but the quark one (the quark plasma). In such  approach the
pion constant $f^*_\pi$ tends to zero, while the mass $m^*_\pi \rightarrow
\infty$ near the point of chiral invariance restoration.  As we discuss in
Sect. 5, this looks unlikely for the real nuclear matter.

 Our approach is based on using the exact pion propagator, renormalized
 in nuclear medium by the insertions of the nucleon-hole and
$\Delta$-hole loops. The short-range correlations are accounted for by
methods of TFFS. The lowest laying singularity in $\rho$ corresponding
to the so-called "pion condensation" is included.  We carried out
 self-consistent calculations with the quark condensate $\kappa(\rho)$
depending on the effective mass $m^*(\rho)$, while the mass $m^*(\rho)$
 itself is determined by (or strongly depends on) $\kappa(\rho)$. Nonlinear
$\rho$-dependence of $\kappa$ is obtained by calculation of the diagram,
shown in the graph of Fig. 1. Calculations include the in-medium values of
nucleon, isobar and pions masses and other parameters ($f^*_\pi, g^*_A,
g'..$). On the other hand, QCD sum rules and NJL-model give the relations
between masses and quark condensates which can be used to determine the
in-medium masses and parameters, if $\kappa(\rho)$ is known. This enabled us
to solve the set of equations (9).

It should be emphasized that, since the dependence of
$m^*(\rho,\kappa(\rho))$ on $\kappa(\rho)$ was treated
self-consistently, the "pion condensation" singularity was pushed out
from the physical sheet. The only effect which can limit the validity
of our calculations at large densities is the accumulation of isobars
 in the ground state of nuclear matter at $\rho\simeq  2.8\rho_0$. In
the general case one should include the possible accumulation of
hyperons. To understand, which of the condensates appears earlier, one
should have better knowledge of hyperon interactions with matter. Analysis
of the problem was started by Pandharipande \cite{13}. This goes beyond the
scope of our paper.

At small densities the nonlinear term $S(\rho)$ diminishes the absolute
value of $\kappa(\rho)$  in comparison with the gas approximation ( the
tendency which was already noted in \cite{11} for very low $\rho)$. In
our self-consistent approach such  behaviour continues up to
$\rho\simeq 2.0\rho_0$.

Taking  relativistic kinematics into account, we have calculated the
expectation values of scalar quark condensate in the symmetric nuclear
matter up to $\rho\simeq 2.8\rho_0$, where  $\kappa$ reaches the value
$\kappa(2.8\rho_0)\simeq 0.1\kappa(0)$. At normal nuclear density $\rho
=\rho_0 =0.17$ fm$^{-3}$, we obtain $\kappa(\rho_0) \simeq 0.6
\kappa(0)$ and the effective nucleon mass $m^*(\rho_0) \simeq 0.6m$. The
latter result is very close to the value used nowadays in QHD \cite{33} to
describe the properties of nuclei.

\subsection*{Acknowledgments} We are grateful to A.A.Anselm,
 I.B.Khriplovich, G.Z.Obrant and M.Rho for useful discussions.
This work was supported by the Russian Fund of Fundamental Research,
Grant No. 96-15-96764

\vspace{1cm}

\subsection{Appendix}

To clarify what happens in the case of  scaling, described by Eq. (40)
we  omit dependence of polarization operator on pion momentum $k$ using
  the mean value of polarization operator $\overline \Pi$  instead. The
quark condensate $|\kappa (\rho)|$ obtained in such a way is plotted in
Fig.  9 as a function of $\overline \Pi$ for two values of density. One
can see the dependence to be a monotonous one.  The curves $A1$ and
$B1$, calculated for $\rho=\rho_A$ and for $\rho=\rho_B > \rho_A$ are
 shown as A1 and B1 in Fig. 9. At some value $\overline \Pi=\Pi_c$,
when the "pion condensation" singularity $\omega_c$ comes the physical
sheet, the function $\kappa(\rho,\Pi_{c}) \rightarrow \infty$. While
both second and third terms of Eq. (2) increase with $\rho$ (for
$S(\rho)$ an extra multiplicative factor $\rho$ comes from the integral
over the nucleon momenta), the function $\kappa(\rho,\overline \Pi)$
becomes steeper for a larger density.  To find self-consistent solution
in this simplified approximation, we have to solve the set of equations

\begin{equation}
\kappa=\kappa(\rho,\overline \Pi) ,
\end{equation}
\begin{equation}
\overline \Pi=\Pi(\rho,\kappa) .
\end{equation}
We show the function $\overline \Pi(\rho,\kappa)$ by  curves $A2$ and
$B2$ in the same figure.  The crossing points of $A1$ and $A2$ (of $B1$
and $B2$) provide the solution.

 It was shown in Sect. 1.2 that the polarization operator can be
approximated as $\overline \Pi(\kappa,\rho)\sim  \rho^{1/3}/\kappa$. So
$\overline \Pi\rightarrow \infty$ at $\kappa\rightarrow 0$ and the mean
value of $\overline \Pi$ increases with $\rho$.

At low densities, there are two solutions: point 1 and point 2, where
the curves $A1$ and $A2$  cross. For $\rho \rightarrow 0$ the first
one ($\kappa_1$ in point 1) matches smoothly with the solution of NJL
gap equation in vacuum: $\kappa_1 \rightarrow \kappa(0)$ when $ \rho
\rightarrow 0$. When the density $\rho$ increases the solutions 1 and 2
draw nearer, merge  with each other (at some $\rho=\rho_{c}$) and go
out into the complex plane.  In this way we lose the real solution at
$\rho >\rho_{c}$.

Assuming that the baryon radius (i.e. parameter $\Lambda$ in the form
factors $d_N$, $d_\Delta$ (see Eq. (14)) follows the same scaling law
 ( $\Lambda^*/\Lambda= f^*_\pi/f_\pi=...$), one obtains more
complicated function $\overline\Pi(\kappa)$ shown in Fig. 9b.  Now for
a very small $\kappa$ the form factors $d_N$,$d_\Delta$ cut off the
integral in the right hand side of Eq. (13) at $k\sim \Lambda << p_F$.
Therefore, the effective polarization operator
$\overline\Pi(\kappa)$ reaches its maximum value and falls down
 when $\kappa \rightarrow 0$, proving us with the third solution
with a very small quark condensate $|\kappa|= \kappa_3$ and thus
with  very small values of hadron masses.  This solution does not
disappear even at very large densities $\rho$.  On the other
hand, for the third solution somewhere at
$\alpha=\kappa_3/\kappa(0)\leq 0.25$ the pion mass

$$m^{*2}_\pi=m^2_\pi\frac{\kappa(0))}{\kappa_3}\propto
\frac {1}{\kappa_3}$$
becomes larger the nucleon mass

$$m^*=m\frac{\kappa_3}{|\kappa(0)|}\propto\kappa_3 ,$$
and we cannot any more apply our
approach, based on the summation of a selected set of Feynman
diagrams corresponding to the lightest pion degrees of freedom.

In any case, at $\rho\geq \rho_{c}$ we lose the primary real solution
1 and face a first order phase transition (mass of a hadron has a
discontinuity) which looks rather strange.


\newpage
\section*{Figure captions}

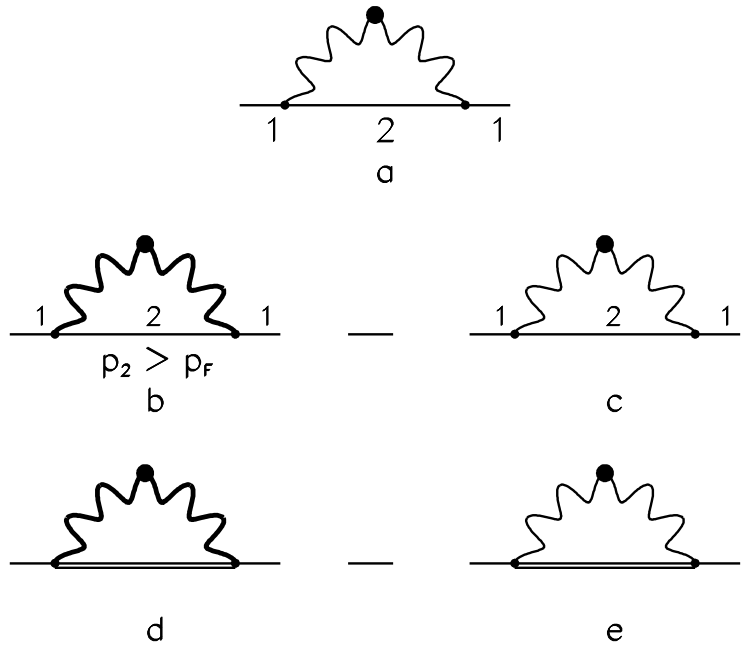
Fig. 1. a) Diagrammatic representation of the interaction of the
operator $\bar qq$ (fat point) with the pion field. Straight
line denotes the nucleon; the wavy line stands for the pion. b,c)~
Diagrammatic expression for  Eq. (13) with  nucleon in the intermediate
state. Thick wavy line denotes the pion propagator renormalized due to
baryon-hole excitations in the framework of TFFS. d,e)~ Diagrammatic
expression for  Eq. (13) with $\Delta$-isobar (double solid line) in the
intermediate state.

\vspace{0.5cm}
Fig. 2. a)~ Full pion propagator  in the medium (thick wavy
line) equal to the sum of the geometrical series of the
nucleon-hole and isobar-hole excitations.  b)~ Polarization
operator of the pion, $\Pi^0_N$, consists of two terms
corresponding to excitation and absorption of the nucleon-hole
pair.  c)~ Pion polarization operator, $\Pi^0_\Delta$, consists of two
terms corresponding to excitation and absorption of the isobar-hole
pair.  Thin wavy line denotes a free pion, solid line with left arrow
 is a hole,  solid line with right arrow denotes a nucleon, double line
 is  $\Delta$-isobar.

\vspace{0.5cm}
Fig. 3. Positions of  singularities of the polarization operator
$\Pi$ in the $\omega$-plane (represented at $p_F=p_{F0}$ and
$k=m_\pi$).  On the right half-plane  $\omega/m_\pi$, in the
interval $\omega/ m_\pi =0.0\div 0.26$, there is the first
logarithmic cut of the $\Pi^0_N$-function, see Eq. (28).  Moving
after
thin arrow from the physical sheet across the first cut, we continue
the movement on the upper logarithmic sheet (thick arrow). In the
interval $\omega/m_\pi =0.26\div 0.45$ there is the second
logarithmic
cut of $\Pi^0_N$, see Eq. (29).  In this case, following the  thin
arrow across the second cut we come to the lower logarithmic sheet
(dashed arrow). Logarithmic cut of the $\Pi^0_\Delta$-function is
located at \\$Re$ $\omega/m_\pi =1.82\div 2.54$ and $Im$ $\omega=
-\Gamma_\Delta/2 =-0.115/2$ GeV.  Since $\Pi^0_N$ and $\Pi^0_\Delta$
are symmetrical in $\omega \leftrightarrow -\omega$ permutation
(Eqs. (19), (20)), there are symmetrical cuts on the left half-plane
$\omega/m_\pi$.

\vspace{0.5cm}
Fig. 4. The condensate branch $\omega_c(k)/m_\pi$. a)~
$\omega_c(k)$ is presented at $p_F=p_{F0}$ for  isobar widths values
$\Gamma_\Delta$= 0.001, 0.01, 0.05, 0.115 GeV (curves 1,2,3,4,
 correspondingly). Solid  lines are disposed on the upper
(unphysical) sheet of the first logarithmic $\Pi^0_N$ cut. Continuation
of this branch to the physical sheet is displayed by dashed curves.
  All curves begin at $k=0$ in the point
$\omega= m^*_\pi=0.80 m_\pi$ (Eq.  (32)).  b)~ $\omega_c(k)$ is
shown at $\Gamma_\Delta$= 0.115 GeV for different values of Fermi
momenta $p_F$=280, 290, 300, 360 MeV (curves 1, 2, 3, 4,
correspondingly).  For curves 3 and 4, the part of $\omega_c(k)$
on the physical sheet is drawn only. The curve 1 for $p_F$=280MeV
is completely on the unphysical sheet.

\vspace{0.5cm}
Fig. 5. The function $S/|\kappa(0)|$. a)~ Dependence of $S/|\kappa(0)|$ on
the variation of nuclear parameters.  Solid curve represents the main result
obtained with $g'_{NN}=1.0$, $g'_{N\Delta}=0.2$, $g'_{\Delta \Delta}=0.8$,
$f_{\pi N\Delta}=2.0$, $\Gamma_\Delta=0.115$ GeV; the other parameters are
described at the end of Sect. 3. Dashed curve corresponds to the calculation
with $f_{\pi N\Delta}=1.7$,  dotted curve to $g'_{\Delta \Delta}$=1.2,
dot-dashed curve to $g'_{NN}=0.7$.  b)~ Dependence of $S/|\kappa(0)|$ on
 the isobar width. Solid curve represents the main result
($\Gamma_\Delta=0.115$ GeV).  Dotted curve corresponds to the calculation
with $\Gamma_\Delta= 0.07$ GeV, dot-dashed curve to $\Gamma_\Delta=0.05$
GeV, dashed curve to $\Gamma_\Delta=0.01$ GeV. c)~Dependence of
$S/|\kappa(0)|$ on the  behaviour type of $m^*(\rho)$.  Solid curve stands
for the  result with  $m^*$ provided by Landau equation (6)
($m^*(\rho=\rho_0)=0.8m$). Dashed curve corresponds to Walecka equation (7)
for $m^*$ ($m^*(\rho=\rho_0)=0.8m$).  Dot-dashed curve is obtained in
framework of Walecka model, with $m^*(\rho= \rho_0)=0.7m$.

\vspace{0.5cm}
Fig. 6. Self-consistent results for the $\bar qq$ expectation value
$\kappa(\rho)/ |\kappa(0)|= <M|\bar qq|M>/ |\kappa(0)|$  (Eq. (2.1))
and $m^*(\rho)/m$ in nuclear matter. a)~ Dependence of
$\kappa(\rho)/|\kappa(0)|$ on the variation of nuclear parameters. Scaling
functions are  $\beta=1, \alpha= \kappa(\rho) /\kappa(0)$. Solid curve is
the main result obtained with $g'_{NN}=1.0$, $g'_{N\Delta}=0.2$, $g'_{\Delta
\Delta}=0.8$, $f_{\pi N\Delta}=2.0$.  Dashed curve corresponds to the
calculation with $f_{\pi N\Delta}$=1.7,  dotted one to $g'_{\Delta
\Delta}=1.2$, dot-dashed curve to $g'_{NN}$=0.7.  b)~ Dependence of
$m^*(\rho)/m$ on the variation of nuclear parameters.  Notation of  curves
are the same as in Fig. 6a. Straight line is drawn for the limit values of
$m^*/m$ (see the end of Sect. 5).  Corresponding reliability limits for the
calculation of $\kappa/\kappa(0)$ are marked by black points.

\vspace{0.5cm}
Fig. 7. Self-consistent results for $\kappa(\rho)/ |\kappa(0)|$
and $m^*(\rho)/m$ in the nuclear matter.  a)~ Dependence of
$\kappa(\rho)/|\kappa(0)$ on the type of scaling functions
$\alpha(\rho), \beta(\rho)$. Solid curve displays the main result
obtained with $\beta$=1, $\alpha$ is calculated by Eq. (37). Long-dashed
curve corresponds to the gas approximation given by Eq. (1). For dotted
curve  $\beta=1, \alpha$ is taken from Eq. (38).  For dot-dashed curve
$\beta=1, \alpha$ is taken from Eq. (39).  Dashed curve corresponds to the
Brown-Rho scaling:  $\beta=1/\alpha, \alpha= \kappa(\rho)/ \kappa(0)$ (40).
b)~ Dependence of $m^*(\rho)/m$ on the kind of scaling functions.  Notations
of the curves are the same as in Fig. 7a. Straight line is drawn for the
limit values of $m^*(\rho)/m$ (see the end of Sect. 5).  Corresponding
reliability limits for the calculation of $\kappa/\kappa(0)$ are marked by
  black points.

\vspace{0.5cm}
Fig. 8. Self-consistent results for $\kappa(\rho)/ |\kappa(0)|$
and $m^*(\rho)/m$ with relativistic corrections.  a)~ Dashed curve  for
  $\kappa(\rho)/ |\kappa(0)|$ is calculated with  relativistic corrections
(Sect. 6) and $\beta=1$, and $\alpha= \kappa/\kappa(0)$,  Eq. (37).  Dotted
curve:  $\beta=1$, and $\alpha$ is taken from Eq. (38). Dot-dashed curve:
$\beta=1$, and $\alpha$ taken from Eq. (39).  Solid curve represents the
main result ($\beta=1, \alpha= \kappa/ \kappa(0)$, Eq. (37))  without
relativistic corrections.  b)~ The results for $m^*(\rho)/m$ calculated with
relativistic corrections.  Notations are the same as in Fig. 8a.  Straight
line restricts  acceptable values of $m^*/m$ from below,
it is obtained using the same method as described above (see the
end of Sect.  5) but with the  isobar polarization operator, Eq. (46).
Corresponding reliability limits for the calculation of
  $\kappa/\kappa(0)$ are marked by  black points.

\vspace{0.5cm}
Fig. 9. The graphical solution of the system of equations (47) and
(48). a)~ The curves $A1$ and $A2$  are  plotted  for the
equations (47) and (48), correspondingly, at certain small
density $\rho_A$.  The crossing points 1 and 2 are the two solutions
of a system of equations. Dashed curves $B1$ and $B2$
correspond to the density value $\rho_B > \rho_A$. There is no solution in
the latter case. b)~ Graphical solution of the set of equations (47) and
(48), when  baryon radius obeys the scaling equation (40). Solid curves $A1$
and $A2$ are  plotted for Eqs. (47) and (48), correspondingly, at small
$\rho_A$. There are three crossing points for solid lines, which are the
three solutions of the system. Curves $B1$ and $B2$ are calculated at
$\rho=\rho_B$, $\rho_B > \rho_A$.  In this case only one solution survives.

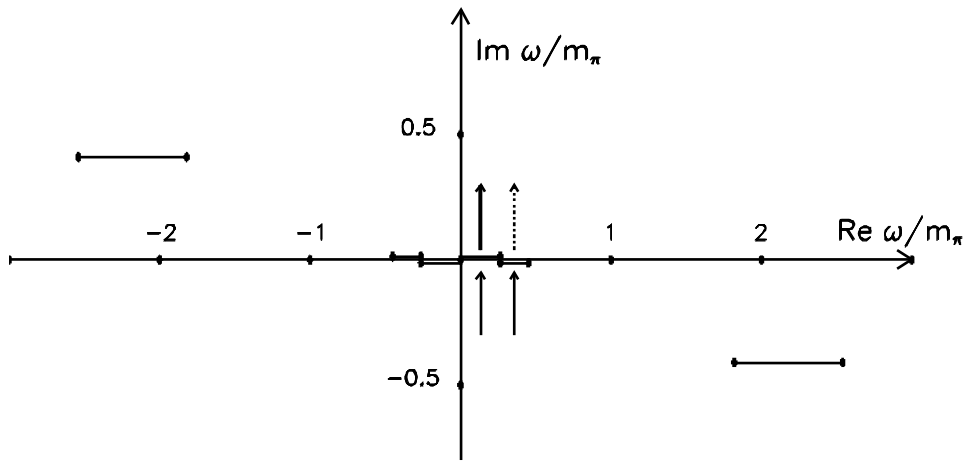
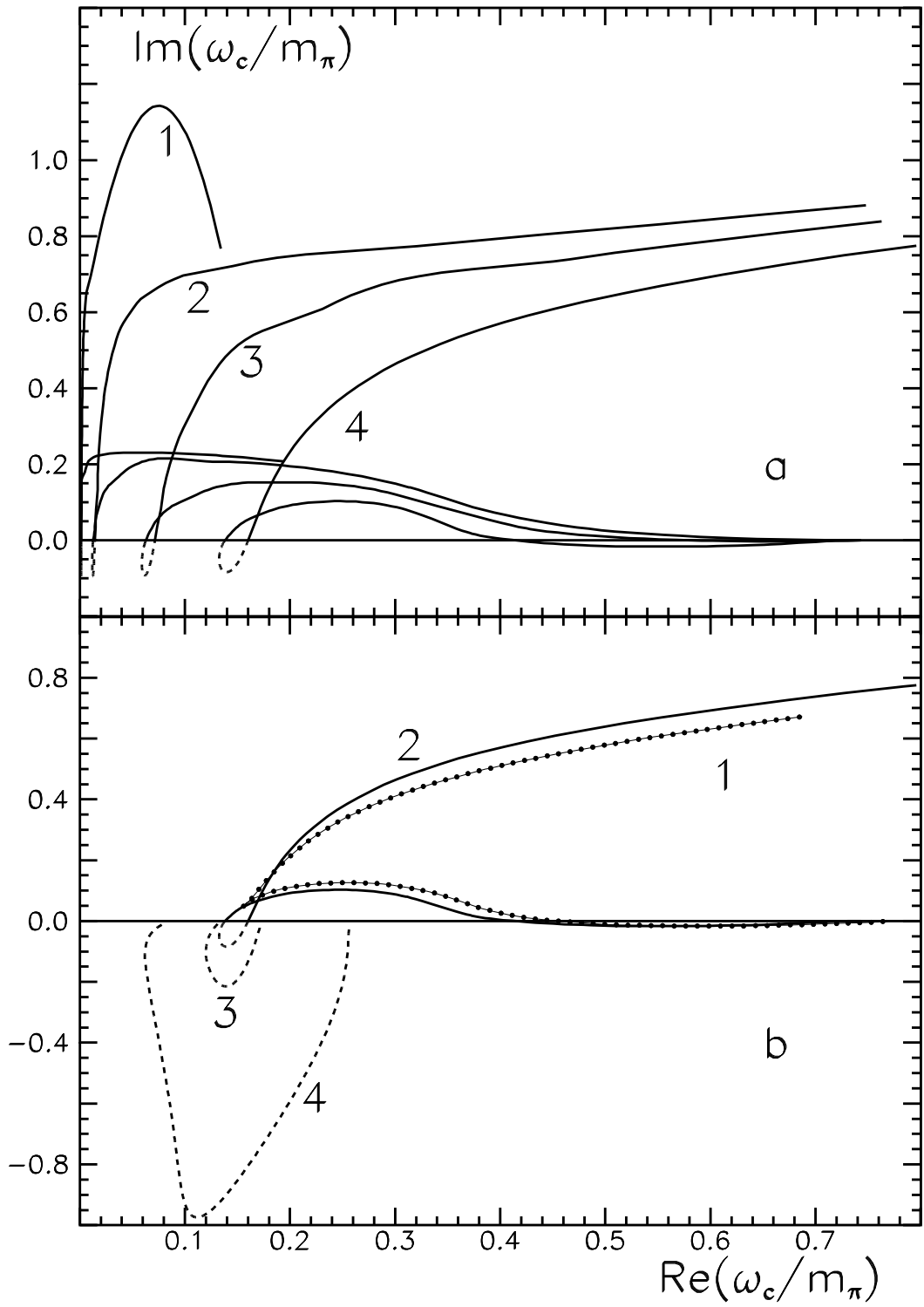
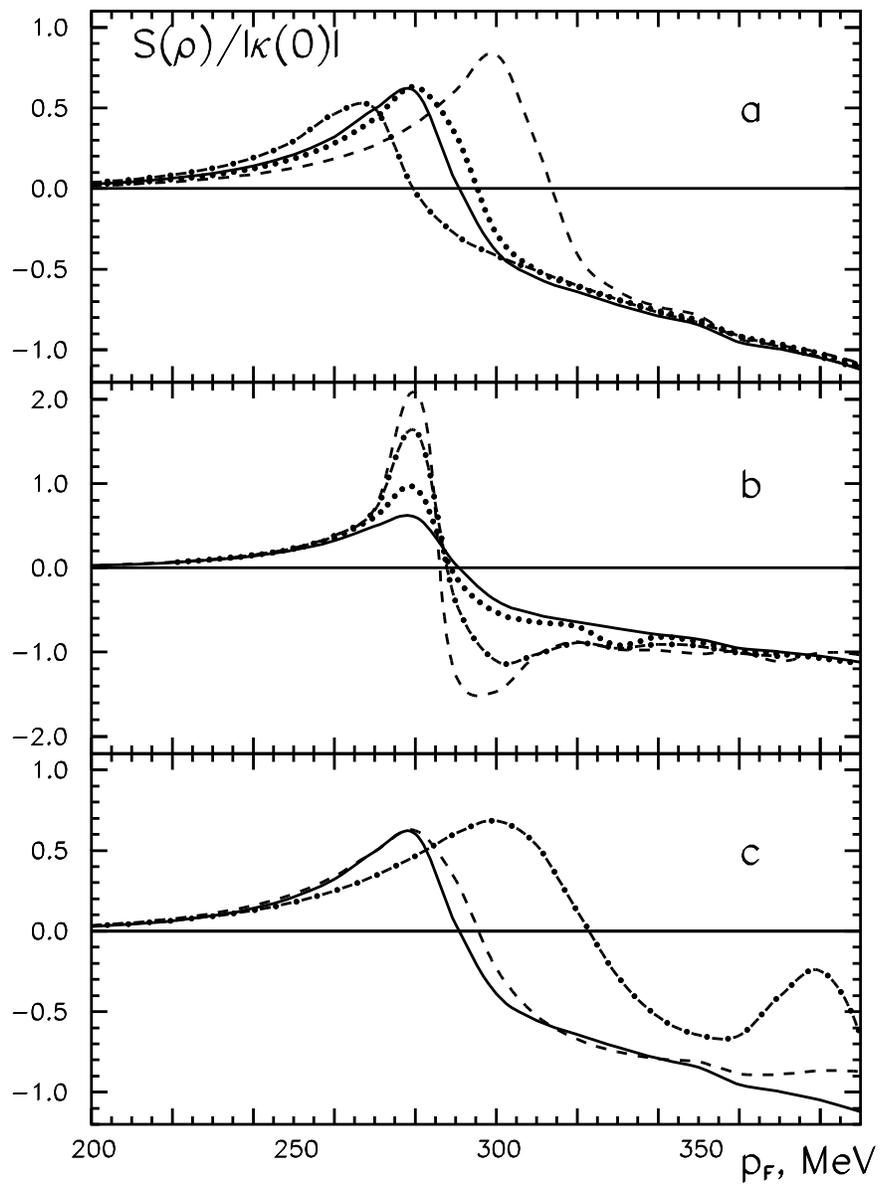
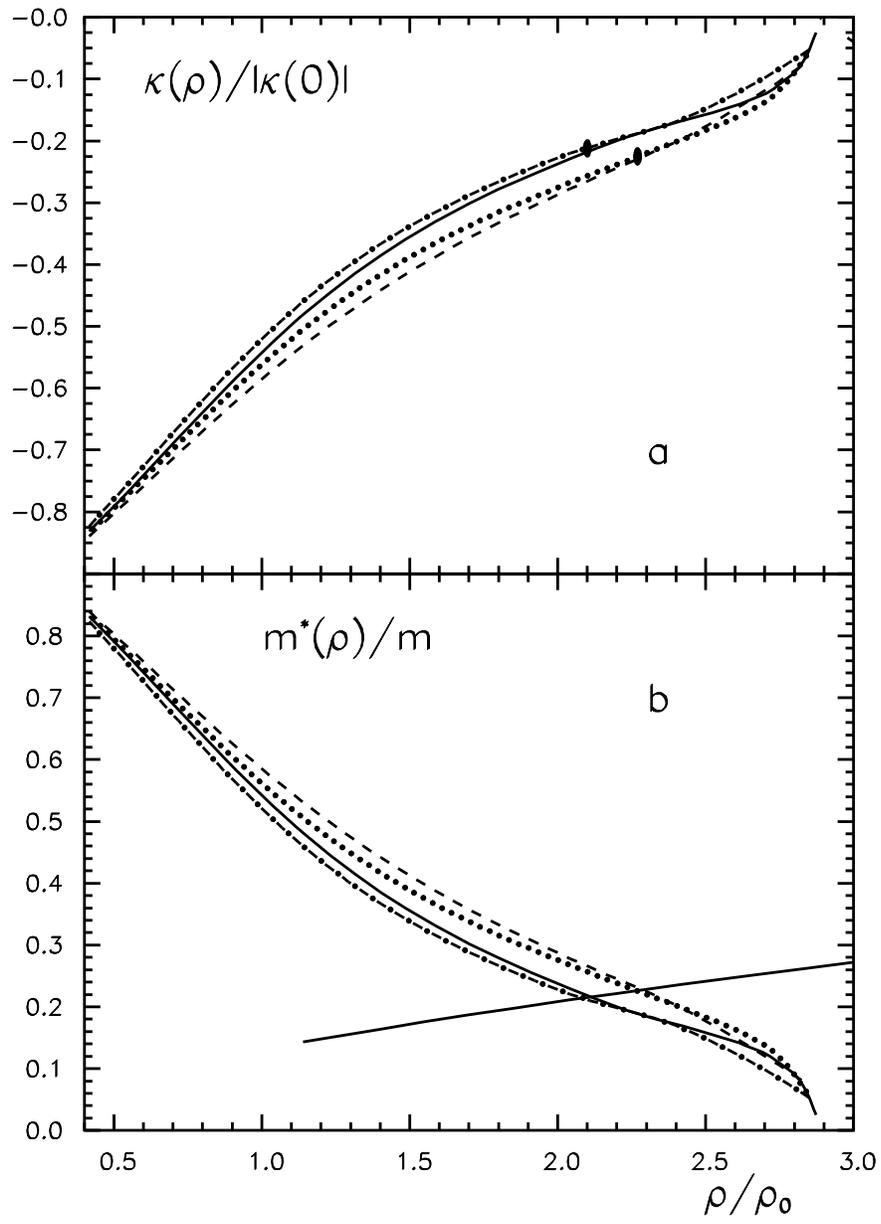
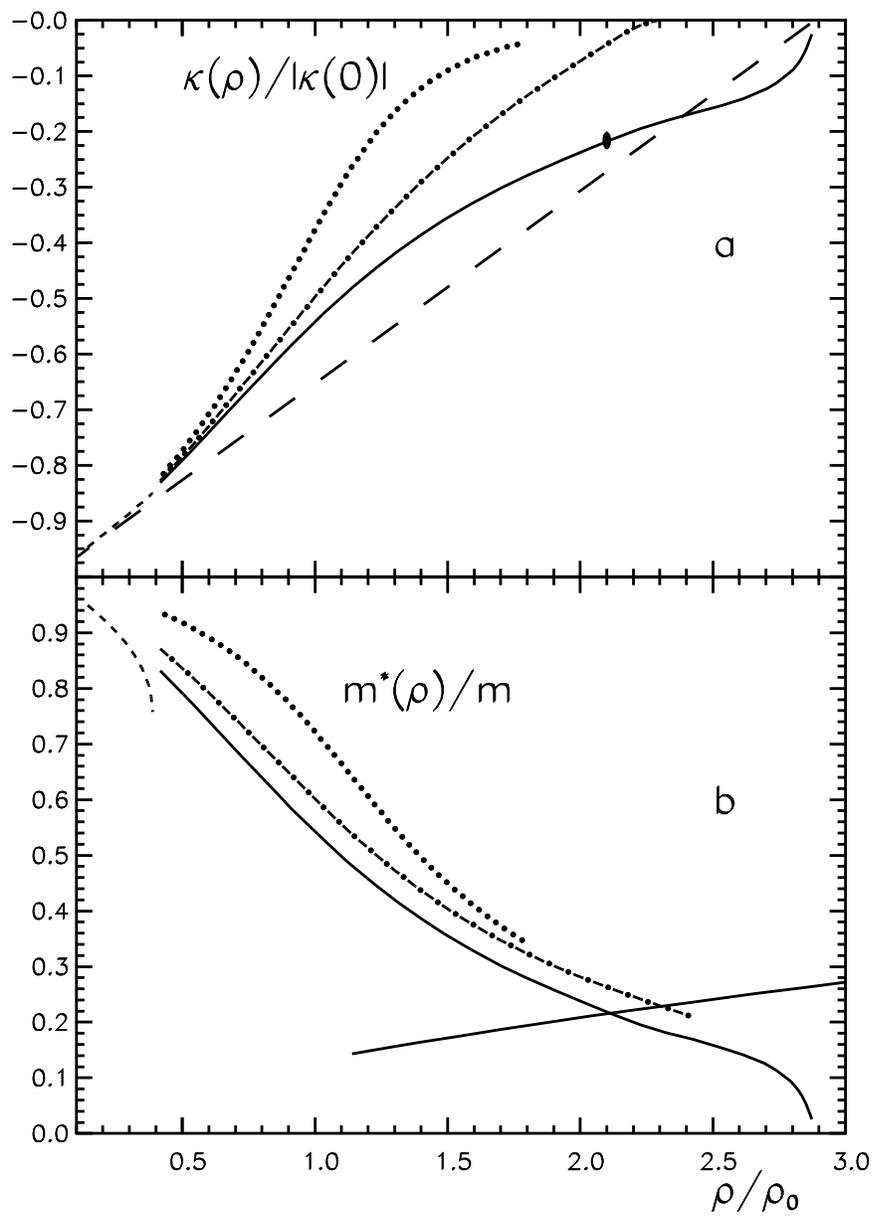
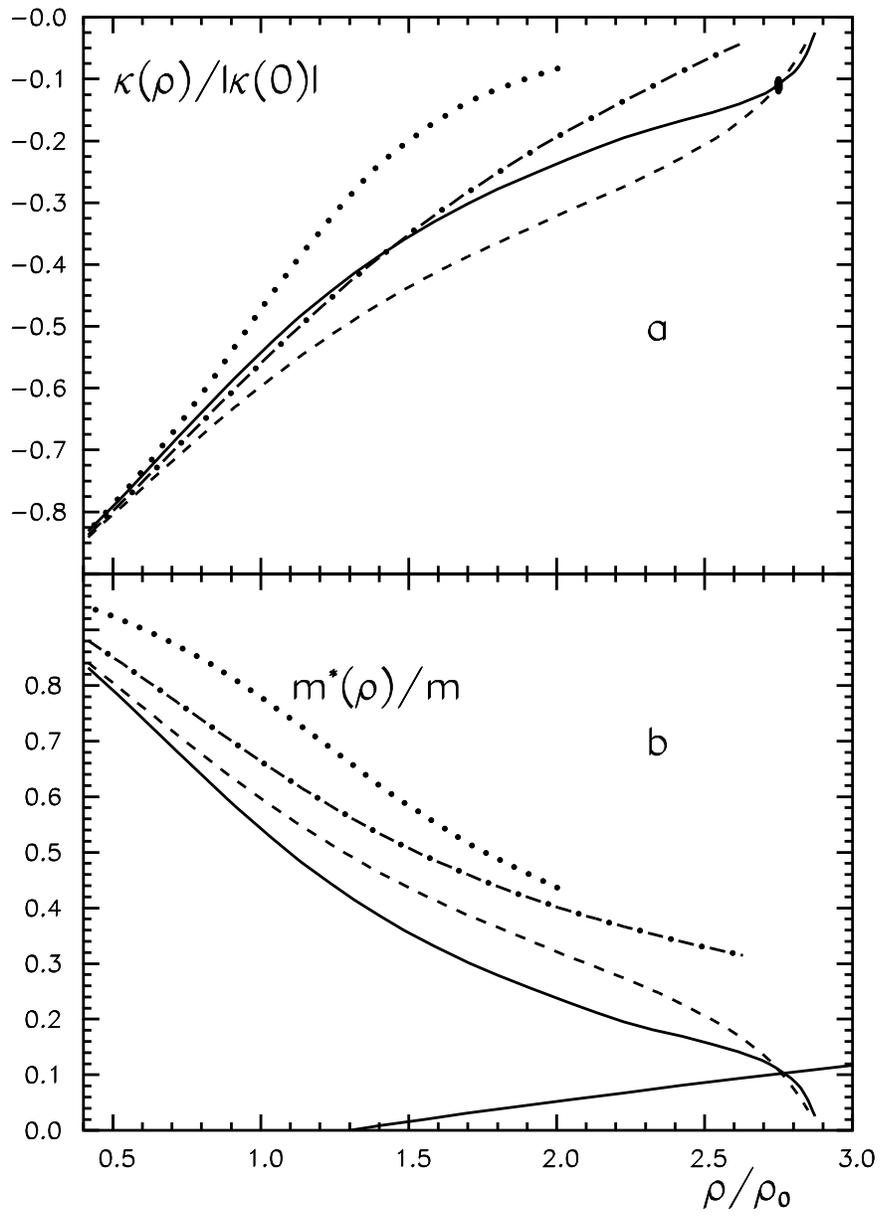
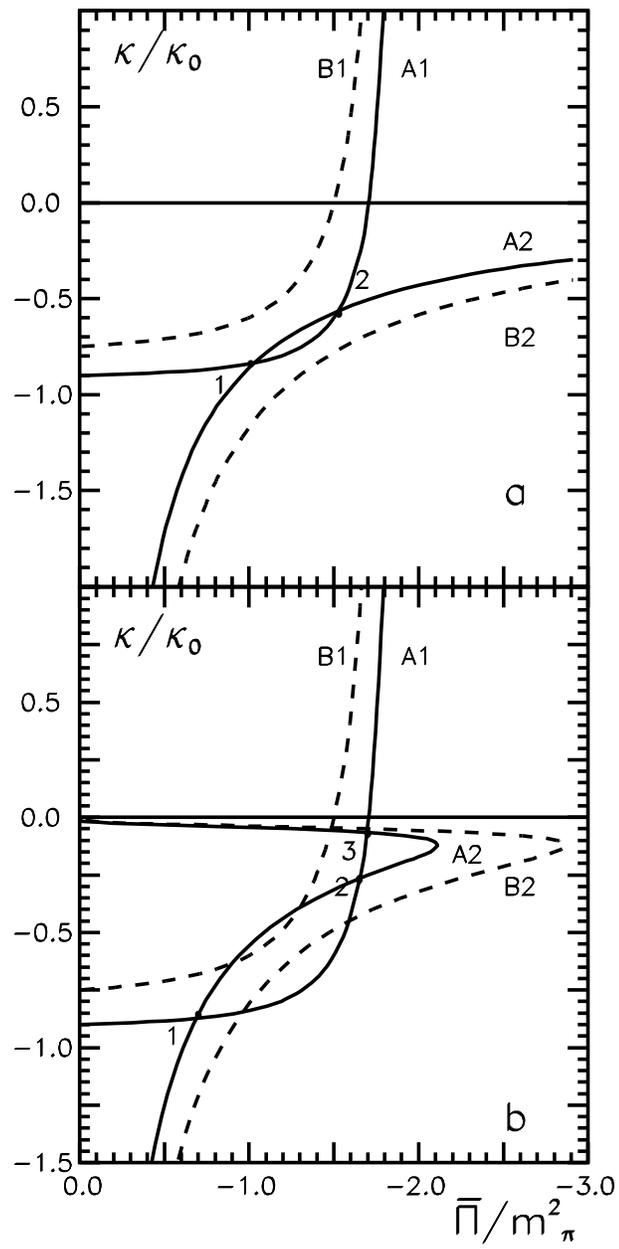
\end{document}